\documentclass[longauth]{aa}  
\usepackage{threeparttable}
\usepackage{graphicx}
\usepackage{txfonts}
\usepackage{gensymb} 
\usepackage{xspace}
\usepackage{tabularx}
\usepackage{xcolor}
\usepackage{placeins}
\usepackage{natbib,twoopt}
\usepackage[breaklinks=true]{hyperref} 
\bibpunct{(}{)}{;}{a}{}{,} 
\makeatletter
\newcommandtwoopt{\citeads}[3][][]{\href{http://adsabs.harvard.edu/abs/#3}%
{\def\hyper@linkstart##1##2{}%
\let\hyper@linkend\@empty\citealp[#1][#2]{#3}}}
\newcommandtwoopt{\citepads}[3][][]{\href{http://adsabs.harvard.edu/abs/#3}%
{\def\hyper@linkstart##1##2{}
\let\hyper@linkend\@empty\citep[#1][#2]{#3}}}
\newcommandtwoopt{\citetads}[3][][]{\href{http://adsabs.harvard.edu/abs/#3}%
{\def\hyper@linkstart##1##2{}
\let\hyper@linkend\@empty\citet[#1][#2]{#3}}}
\newcommandtwoopt{\citeyearads}[3][][]%
{\href{http://adsabs.harvard.edu/abs/#3}
{\def\hyper@linkstart##1##2{}%
\let\hyper@linkend\@empty\citeyear[#1][#2]{#3}}}
\makeatother
\def\ms{\hbox{m\,s$^{-1}$}}         
\def\m2s2{\hbox{\,m$^{2}$\,s$^{-2}$}} 
\def\vsini{\hbox{$v$\,sin\,$i_{\star}$}}      
\def\Msun{$M_{\odot}$\xspace}             
\def\Rsun{$R_{\odot}$\xspace}
\def\Mjup{\hbox{$\mathrm{M}_{\rm J}$}}
\def\Rjup{\hbox{$\mathrm{R}_{\rm J}$}}

\def\ten[#1]{$\;\times 10^{#1}$}

\def\logg{$\log g$}

\newcommand{\Rnom}{\hbox{$\mathcal{R}^{\rm N}_{\odot}$}} 

\newcommand{\GMnom}{\hbox{$\mathcal{(GM)}^{\rm N}_{\odot}$}}

\newcommand{\Renom}{\hbox{$\mathcal{R}^{\rm N}_{e \rm E}$}}

\newcommand{\GMenom}{\hbox{$\mathcal{(GM)}^{\rm N}_{\rm E}$}}

\newcommand{\RJnom}{\hbox{$\mathcal{R}^{\rm N}_{e \rm J}$}}

\newcommand{\GMJnom}{\hbox{$\mathcal{(GM)}^{\rm N}_{\rm J}$}}

\newcommand{\rebound}{{\sc \tt REBOUND}\xspace}

\newcommand{\juliet}{{\sc \tt juliet}\xspace}
\newcommand{\batman}{{\sc \tt batman}\xspace}
\newcommand{\radvel}{{\sc \tt radvel}\xspace}
\newcommand{\celerite}{{\sc \tt celerite}\xspace}
\newcommand{\george}{{\sc \tt george}\xspace}
\newcommand{\dynesty}{{\sc \tt dynesty}\xspace}

\newcommand{\specmatch}{{\sc \tt SpecMatch-Emp}\xspace}
\newcommand{\starry}{{\sc \tt starry}\xspace}

\newcommand{\drs}{{\sc \tt DRS; v. 3.0.0}\xspace}

\newcommand{\MEarth}{$\mathrm{M_E}$\xspace}

\def\logg{$\log g$}
\def\Msun{$M_{\odot}$\xspace}            
\def\Rsun{$R_{\odot}$\xspace}
\newcommand{\orcid}[1]{\protect\href{https://orcid.org/#1}{\protect\includegraphics[width=8pt]{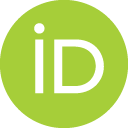}}}
\hypersetup{colorlinks=true, citecolor=blue, linkcolor=blue}

\begin{document} 

\title{TOI-4860\,b, a short-period giant planet transiting an M3.5 dwarf}
   
   \author{
J.M.~Almenara\orcid{0000-0003-3208-9815}\inst{\ref{grenoble},\ref{geneva}}
        \and X.~Bonfils\orcid{0000-0001-9003-8894}\inst{\ref{grenoble}}
        \and E.~M.~Bryant\orcid{0000-0001-7904-4441}\inst{\ref{mssl},\ref{warwick1}}
        \and A.~Jord\'an\orcid{0000-0002-5389-3944}\inst{\ref{uai},\ref{MAS},\ref{DO}}
        \and G.~H\'ebrard\inst{\ref{iap},\ref{ohp}}
        \and E.~Martioli\orcid{0000-0002-5084-168X}\inst{\ref{lna},\ref{iap}}
        \and A.~C.~M.~Correia\orcid{0000-0002-8946-8579}\inst{\ref{coimbra},\ref{imcce}}
        \and N.~Astudillo-Defru\orcid{0000-0002-8462-515X}\inst{\ref{concepcion}}
        \and C.~Cadieux\orcid{0000-0001-9291-5555}\inst{\ref{montreal2}}
        \and L.~Arnold\orcid{0000-0002-0111-1234}\inst{\ref{cfht}}
        \and \'E.~Artigau\orcid{0000-0003-3506-5667}\inst{\ref{montreal2}}
        \and G.\'A.~Bakos\inst{\ref{princeton}}
        \and S.C.C.~Barros\inst{\ref{porto1}, \ref{porto2}}
        \and D.~Bayliss\inst{\ref{warwick1}}
        \and F.~Bouchy\inst{\ref{geneva}}
        \and G.~Bou\'{e}\inst{\ref{imcce}}
        \and R.~Brahm\orcid{0000-0002-9158-7315}\inst{\ref{uai},\ref{MAS},\ref{DO}}
        \and A.~Carmona\inst{\ref{grenoble}}
        \and D.~Charbonneau\inst{\ref{harvard}}
        \and D.R.~Ciardi\orcid{0000-0002-5741-3047}\inst{\ref{pasadena}}
        \and R.~Cloutier\orcid{0000-0001-5383-9393}\inst{\ref{mcmaster}}
        \and M.~Cointepas\inst{\ref{grenoble}, \ref{geneva}}
        \and N.J.~Cook\orcid{0000-0003-4166-4121}\inst{\ref{montreal2}}
        \and N.B.~Cowan\inst{\ref{quebec1}, \ref{quebec2}}
        \and X.~Delfosse\inst{\ref{grenoble}}
        \and J.~Dias~do~Nascimento\orcid{0000-0001-7804-2145}\inst{\ref{harvard}, \ref{natal}} 
        \and J.-F.~Donati\orcid{0000-0001-5541-2887}\inst{\ref{toulouse}}
        \and R.~Doyon\inst{\ref{montreal2}}
        \and T.~Forveille\orcid{0000-0003-0536-4607}\inst{\ref{grenoble}}
        \and P.~Fouqu\'e\orcid{0000-0002-1436-7351}\inst{\ref{toulouse}}
        \and E.~Gaidos\orcid{0000-0002-5258-6846}\inst{\ref{hawaii}}
        \and E.A.~Gilbert\orcid{0000-0002-0388-8004}\inst{\ref{jpl}}
        \and J.~Gomes~da~Silva\inst{\ref{porto1}}
        \and J.D.~Hartman\inst{\ref{princeton}}
        \and K.~Hesse\inst{\ref{middletown}}
        \and M.J.~Hobson\orcid{0000-0002-5945-7975}\inst{\ref{mpia}, \ref{MAS}}
        \and J.M.~Jenkins\inst{\ref{ames}}
        \and F.~Kiefer\inst{\ref{lesia}}
        \and V.B.~Kostov\inst{\ref{greenbelt}, \ref{seti}}
        \and J.~Laskar\inst{\ref{imcce}}
        \and M.~Lendl\inst{\ref{geneva}}
        \and A.~L'Heureux\orcid{0009-0005-6135-6769}\inst{\ref{montreal2}}
        \and J.H.C.~Martins\inst{\ref{porto1}}
        \and K.~Menou\inst{\ref{toronto1}, \ref{toronto2}, \ref{toronto3}}
        \and C.~Moutou\inst{\ref{toulouse}}
        \and F.~Murgas\orcid{0000-0001-9087-1245}\inst{\ref{iac}, \ref{ull}}
        \and A.S.~Polanski\orcid{0000-0001-7047-8681}\inst{\ref{lawrence}, \ref{pasadena}}
        \and D.~Rapetti\orcid{0000-0003-2196-6675}\inst{\ref{ames},\ref{USRA}} 
        \and E.~Sedaghati\orcid{0000-0002-7444-5315}\inst{\ref{ESO}}
        \and H.~Shang\inst{\ref{taipei}}
}
      \institute{
        Univ. Grenoble Alpes, CNRS, IPAG, F-38000 Grenoble, France\label{grenoble}
        \and Observatoire de Gen\`eve, Département d’Astronomie, Universit\'e de Gen\`eve, Chemin Pegasi 51b, 1290 Versoix, Switzerland\label{geneva}
        \and Mullard Space Science Laboratory, University College London, Holmbury St Mary, Dorking, Surrey, RH5 6NT, UK\label{mssl}
        \and Department of Physics, University of Warwick, Coventry, UK\label{warwick1}
        \and Facultad de Ingenier\'ia y Ciencias, Universidad Adolfo Ib\'a\~nez, Av.\ Diagonal las Torres 2640, Pe\~nalol\'en, Santiago, Chile\label{uai}
        \and Millennium Institute for Astrophysics (MAS), Santiago, Chile\label{MAS}
        \and Data Observatory Foundation, Chile\label{DO}
        \and Institut d'astrophysique de Paris, UMR7095 CNRS, Universit\'e Pierre \& Marie Curie, 98bis boulevard Arago, 75014 Paris, France\label{iap}
        \and Observatoire de Haute-Provence, CNRS, Universit\'e d'Aix-Marseille, 04870 Saint-Michel-l'Observatoire, France\label{ohp}
        \and Laborat\'{o}rio Nacional de Astrof\'{i}sica, Rua Estados Unidos 154, 37504-364, Itajub\'{a} - MG, Brazil\label{lna}
        \and CFisUC, Departamento de F\'isica, Universidade de Coimbra, 3004-516 Coimbra, Portugal\label{coimbra}
        \and IMCCE, UMR8028 CNRS, Observatoire de Paris, PSL University, Sorbonne Univ., 77 av. Denfert-Rochereau, 75014 Paris, France\label{imcce}
        \and Departamento de Matem\'{a}tica y F\'{i}sica Aplicadas, Universidad Cat\'{o}lica de la Sant\'{i}sima Concepci\'{o}n, Alonso de Rivera 2850, Concepci\'{o}n, Chile\label{concepcion}
        \and Universit\'e de Montr\'eal, D\'epartement de Physique \& Institut Trottier de Recherche sur les Exoplan\`etes, Montr\'eal, QC H3C 3J7, Canada\label{montreal2}
        \and Canada-France-Hawaii Telescope (CFHT) Corporation, CNRS UAR2208, 65-1238 Mamalahoa Hwy, Kamuela, HI 96743, USA\label{cfht}
        \and Department of Astrophysical Sciences, Princeton University, NJ 08544, USA\label{princeton}
        \and Instituto de Astrof\'isica e Ci\^encias do Espa\c{c}o, Universidade do Porto, CAUP, Rua das Estrelas, 4150-762 Porto, Portugal\label{porto1}
        \and Departamento de Fisica e Astronomia, Faculdade de Ciencias, Universidade do Porto, Rua Campo Alegre, 4169-007 Porto, Portugal\label{porto2}
        \and Center for Astrophysics \textbar \ Harvard \& Smithsonian, 60 Garden Street, Cambridge, MA 02138, USA\label{harvard}
        \and NASA Exoplanet Science Institute – Caltech/IPAC, 1200 E. California Blvd, Pasadena, CA 91125 USA\label{pasadena}
        \and Department of Physics \& Astronomy, McMaster University, 1280 Main St W, Hamilton, ON, L8S 4L8, Canada\label{mcmaster}
        \and Department of Physics, McGill University, 3600 rue University, Montr\'eal, Qu\'ebec, H3A 2T8, Canada\label{quebec1}
        \and Department of Earth and Planetary Sciences, McGill University, 3450 rue University, Montr\'eal, Qu\'ebec, H3A 0E8, Canada\label{quebec2}
        \and Univ. Federal do Rio G. do Norte, UFRN, Dep. de F\'isica, CP 1641, 59072-970, Natal, RN, Brazil\label{natal}
        \and Universit\'e de Toulouse, CNRS, IRAP, 14 avenue Belin, 31400 Toulouse, France\label{toulouse}
        \and Dept. of Earth Sciences, University of Hawai'i at M\"{a}noa, Honolulu, HI 96822 USA\label{hawaii}
        \and Jet Propulsion Laboratory, California Institute of Technology, 4800 Oak Grove Drive, Pasadena, CA 91109, USA\label{jpl}
        \and Wesleyan University, Middletown, CT 06459, USA\label{middletown}
        \and Max Planck Institute for Astronomy, K{\"{o}}nigstuhl 17, 69117 - Heidelberg, Germany\label{mpia}
        \and NASA Ames Research Center, Moffett Field, CA 94035, USA\label{ames}
        \and LESIA, Observatoire de Paris, Universit\'e PSL, CNRS, Sorbonne Universit\'e, Universit\'e Paris Cit\'e, 5 place Jules Janssen, 92195 Meudon, France\label{lesia}
        \and NASA Goddard Space Flight Center, 8800 Greenbelt Road, Greenbelt, MD 20771, USA\label{greenbelt}
        \and SETI Institute, 189 Bernardo Ave, Suite 200, Mountain View, CA 94043, USA\label{seti}
        \and Physics \& Astrophysics Group, Dept. of Physical \& Environmental Sciences, University of Toronto Scarborough, 1265 Military Trail, Toronto, Ontario, M1C 1A4, Canada\label{toronto1}
        \and David A. Dunlap Department of Astronomy \& Astrophysics, University of Toronto. 50 St. George Street, Toronto, Ontario, M5S 3H4, Canada\label{toronto2}
        \and Department of Physics, University of Toronto, 60 St George Street, Toronto, Ontario, M5S 1A7, Canada\label{toronto3}
        \and Instituto de Astrofísica de Canarias (IAC), E-38200 La Laguna, Tenerife, Spain\label{iac}
        \and Dept. Astrofísica, Universidad de La Laguna (ULL), E-38206 La Laguna, Tenerife, Spain\label{ull}
        \and Department of Physics and Astronomy, University of Kansas, 1251 Wescoe Hall Drive, Lawrence, KS 66045, USA\label{lawrence}
        \and Research Institute for Advanced Computer Science, Universities Space Research Association, Washington, DC 20024, USA\label{USRA}
        \and European Southern Observatory, Alonso de C\'ordova 3107, Vitacura, Regi\'on Metropolitana, Chile\label{ESO}
        \and Institute of Astronomy and Astrophysics, Academia Sinica, Taipei 10617, Taiwan\label{taipei}
}

\date{Received ; Accepted}

\abstract
{
We report the discovery and characterisation of a giant transiting planet orbiting a nearby M3.5V dwarf (d = 80.4\,pc, $G$ = 15.1\,mag, $K$=11.2\,mag, $R_\star = 0.358\pm0.015$~\Rsun, $M_\star = 0.340\pm0.009$~\Msun). Using the photometric time series from TESS sectors 10, 36, 46, and 63 and near-infrared spectrophotometry from ExTrA, we measured a planetary radius of $0.77 \pm 0.03~\Rjup$ and an orbital period of 1.52 days. With high-resolution spectroscopy taken by the CFHT/SPIRou and ESO/ESPRESSO spectrographs, we refined the host star parameters ($\rm [Fe/H] = 0.27\pm0.12$) and measured the mass of the planet ($0.273 \pm 0.006$~\Mjup). Based on these measurements, TOI-4860\,b joins the small set of massive planets ($>80$~\MEarth) found around mid to late M dwarfs ($<0.4$~\Rsun), providing both an interesting challenge to planet formation theory and a favourable target for further atmospheric studies with transmission spectroscopy. We identified an additional signal in the radial velocity data that we attribute to an eccentric planet candidate ($e=0.66\pm0.09$) with an orbital period of $427\pm7$~days and a minimum mass of $1.66\pm 0.26$~\Mjup, but additional data would be needed to confirm this.
}


\maketitle

\section{Introduction}
Twenty-five years ago, the discovery of GJ~876\,b, a planet with twice the mass of Jupiter, marked an important milestone in exoplanet research \citep{marcy1998,delfosse1998}. It was the first planet detected around an M dwarf, and only the ninth exoplanet found around a main-sequence star at that time. The discovery of GJ~876\,b soon after the detection of 51~Pegasi\,b around a solar-type star \citep{mayor1995} suggested that giant planets could form just as easily around late-type stars. The pace of planet detections around M dwarfs, however, remained much behind that around FGK stars, and giant planets appear to have a decreasing occurrence rate with decreasing stellar mass \citep{endl2006,johnson2007,bonfils2005}.

Probing the whole sky for transiting planets, the Transiting Exoplanet Survey Satellite \citep[TESS;][]{ricker2015} survey vastly expanded the target list and therefore was better able to detect rare outcomes of planet formation. A new planet population emerged: giant planets around very low-mass stars (e.g. TOI-519\,b, \citealt{kagetani2023}; TOI-3714\,b, \citealt{canas2022}; TOI-3757\,b, \citealt{kanodia2022}; TOI-5205\,b, \citealt{kanodia2023}; and TOI-3235\,b, \citealt{hobson2023}). These planets are massive in comparison to expectations for protoplanetary disk masses around M dwarfs \citep{andrews2013,gaidos2017} and therefore are interesting for the investigation of planetary formation. 

Recently, \citet{bryant2023} performed a systematic search in the TESS photometry for giant planets transiting M dwarfs and measured an occurrence rate of $0.194\pm0.072\%$ giant planets with periods below 10~days per M dwarf (0.08$-$0.71~\Msun), and $0.134\pm0.069\%$ for stellar masses between 0.088 and 0.4~\Msun. \citet{gan2023}, also using the TESS data, measured a hot Jupiter occurrence rate around early-type M dwarfs (0.45$-$0.65~\Msun) of $0.27\pm0.09\%$. For comparison, the hot Jupiter occurrence rate around G-type stars is $0.55\pm0.14\%$ \citep{beleznay2022}. Further characterisation of these planets' masses, orbits, and general architecture is now key to understanding how they form.

We report here the confirmation and mass measurement of TOI-4860\,b\footnote{We would like to acknowledge that another team has also conducted work on TOI-4860, utilising the same TESS data. Both teams made a mutual decision to submit our papers simultaneously, without discussing our respective findings. Therefore, both the current paper and the one by \citet{triaud2023} present completely independent analyses of TOI-4860.}. Ground-based photometry confirms that the transits around TOI-4860, initially identified by TESS photometry, are produced by a giant planet. Doppler spectroscopy measures its mass and helps characterise the host star. This new planet detection adds to a small set of giant planets detected around stars with $M_\star<0.4$~\Msun, a stellar mass regime where standard core accretion models are not able to form giants \citep[e.g.][]{burn2021}, and therefore provides a challenge to the hypothesis underlying these models. Additionally, TOI-4860\,b is an interesting target for atmospheric characterisation with transmission spectroscopy.

In Sect.~\ref{section:observations} we describe the data used to detect and characterise TOI-4860\,b. In Sect.~\ref{section:stellarparameters} we characterise its host star, and in Sect.~\ref{section.analysis} we model the radial velocity (RV) and transit data. Finally, in Sect.~\ref{section.results} we discuss the results of our work.

\section{Observations}\label{section:observations}
\subsection{TESS photometry}

TOI-4860 was observed in four TESS sectors: one sector during the primary mission (sector 10, March and April 2019), two sectors during the first extended mission (sector 36, March and April 2021; sector 46, December 2021), and one sector in the second extended mission (sector 63, March and April 2023). Only full frame image (FFI) data were collected for sectors 10, 36, and 46, at respectively 30~min, 10~min, and 10~min cadence. For sector~63, data were obtained as both postage stamps at a 2-minute cadence and 200-second FFIs.

A transit signature with a 1.52-day period and 6\% depth was first identified from the FFIs by the Faint Star Search \citep{kunimoto2022} using data products from the Quick-Look Pipeline \citep[QLP;][]{huang2020a,huang2020b,kunimoto2021}. The TESS Science Processing Operations Center \citep[SPOC;][]{jenkins2016,caldwell2020} at NASA Ames Research Center conducted a transit search of Sector 36 on 5 May 2021 on an FFI 10-minute cadence observation with an adaptive, noise-compensating matched filter \citep{jenkins2002,jenkins2010,jenkins2020}, producing a threshold crossing event for which an initial limb-darkened transit model was fitted \citep{li2019} and a suite of diagnostic tests were conducted to help make or break the planetary nature of the signal \citep{twicken2018}. The signature and associated diagnostics were reviewed by the TESS Science Office and TOI-4860.01 was alerted as a TESS Object of Interest (TOI) on 21 December 2021 \citep{guerrero2021}. The signal was repeatedly recovered as additional observations were made in Sectors 46 and 63, and the transit signature passed all the diagnostic tests presented in the data validation reports. The difference image centroiding analyses constrain the location of the host star to within $3.5\pm2.5\arcsec$ of the source of the transit signal.

\begin{figure*}[h]
    \centering
    \includegraphics[width=1.0\textwidth]{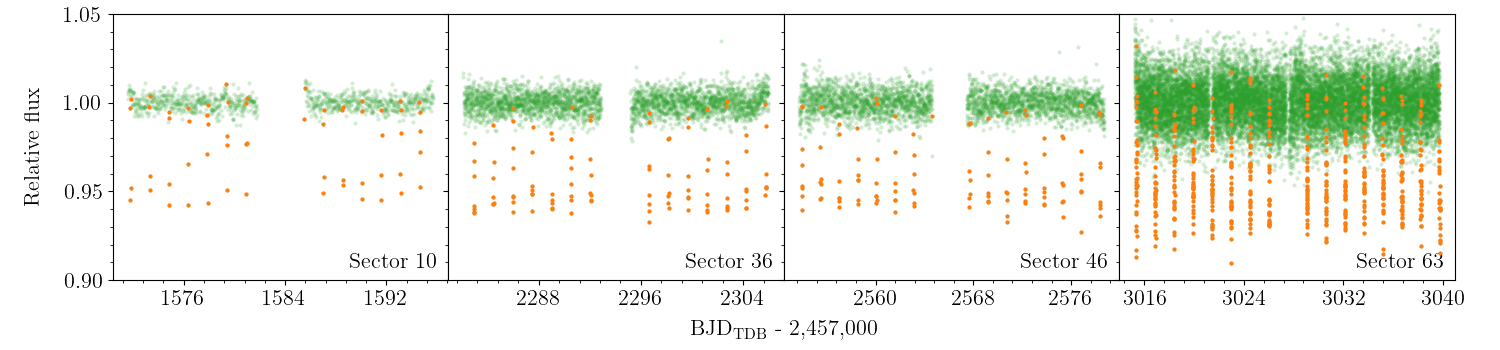}
    \caption{TESS PDCSAP photometry time series (sectors 10, 36, 46, and 63; green dots), with the transits of TOI-4860\,b highlighted in orange. The sampling is 30~min for sector 10, 10~min for sectors 36 and 46, and 2~min for sector 63.}
    \label{fig:tess}
\end{figure*}

\begin{figure}[h]
   \centering
   \includegraphics[width=0.47\textwidth]{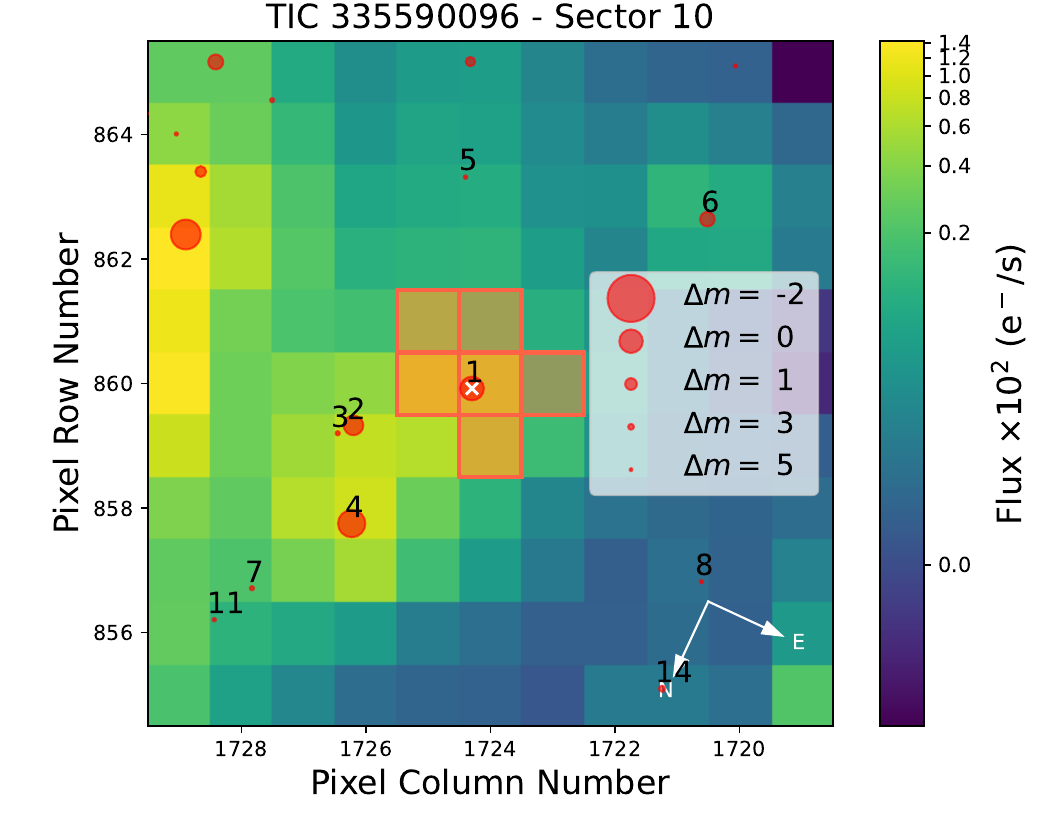}
      \caption{TESS TPF image of TOI-4860 in Sector 10 \citep[created with {\sc \tt tpfplotter};][]{aller2020}. The electron counts are colour-coded. The red-bordered pixels are used in the SAP. The size of the red circles indicates the \textit{Gaia} Data Release 2 magnitudes of all nearby stars \citep{gaia2018}.}
      \label{fig:tpf}
\end{figure}

For the analysis in Sect.~\ref{section.analysis}, we used the Presearch Data Conditioning Simple Aperture Photometry (PDCSAP; \citealt{smith2012}, \citealt{stumpe2012,stumpe2014}, \citealt{caldwell2020}) light curve of TOI-4860 (Fig.~\ref{fig:tess}), produced by the SPOC, which is corrected for dilution in the TESS aperture by known contaminating sources. This is critical to ensure that no visually close-by targets that could affect the depth of the transit are present in the 21\arcsec\ TESS pixel and to check for a contaminating eclipsing binary. Figure~\ref{fig:tpf} shows a plot of the target pixel file (TPF) from sector 10 and the aperture mask that is used for the simple aperture photometry (SAP). We can see that no star overlaps the TESS aperture down to six~magnitudes fainter than TOI-4860.

\subsection{Inspection for contaminants}

\subsubsection{Proper motion}

\begin{figure*}[h]
   \centering
   \includegraphics[width=\textwidth]{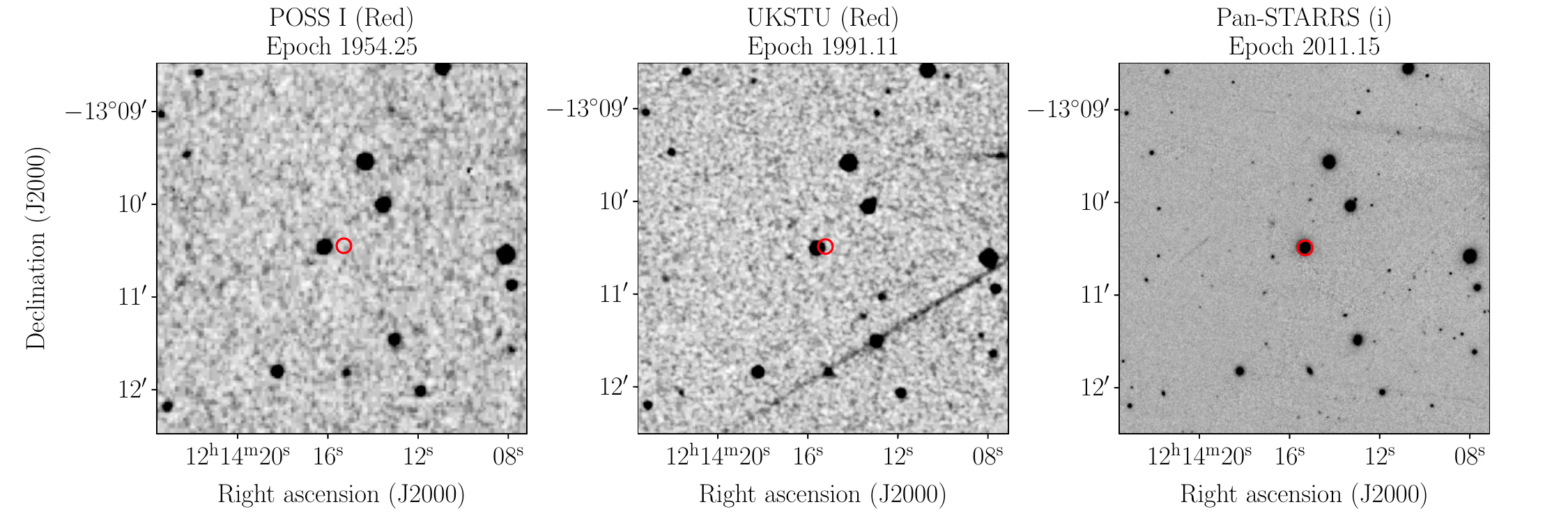}
      \caption{Images extracted from the Digitized Sky Survey (DSS; retrieved from \url{https://archive.stsci.edu/cgi-bin/dss_form}) and the Pan-STARRS survey \citep{chambers2016}. From left to right, the epochs are years 1954, 1991, and 2011. The open red circle marks the position of the star as determined by \textit{Gaia} \citep{gaiaDR3}. There is no background star at the current position of TOI-4860.}
      \label{fig:pm}
\end{figure*}

TOI-4860 has a proper motion of 177 mas/yr \citep{gaiaDR3}, and therefore has moved by 12\arcsec\ from its position on the Palomar Observatory Sky Survey \citep[POSS;][]{minkowski1963,lasker1996} I red plate taken in 1954. This offset is large enough to check for background stars at the current position of TOI-4860, and we see none (Fig.~\ref{fig:pm}).

\subsubsection{\textit{Gaia} assessment}
In addition to proper motion, we used \textit{Gaia} \citep{gaia2016, gaiaDR3} to identify any wide stellar companions that may be bound members of the system. Typically, these stars are already in the TESS Input Catalog \citep{stassun2019} and their flux dilution to the transit has already been accounted for in the transit fits and associated derived parameters. There are no additional widely separated companions identified by \textit{Gaia} that have the same distance and proper motion as TOI-4860 \citep[see also][]{mugrauer2020,mugrauer2021}.

The \textit{Gaia} Data Release 3 (DR3) astrometry provides additional information on the possibility of inner companions that may have gone undetected by either \textit{Gaia} or high-resolution imaging. The \textit{Gaia} re-normalised unit weight error (RUWE) is a metric similar to a reduced chi-square, and values $\lesssim 1.4$ indicate that the \textit{Gaia} astrometric solution is consistent with the star being single, whereas RUWE values $\gtrsim 1.4$ indicate excess astrometric noise, which is often caused by an unseen companion \citep[e.g.][]{ziegler2020}. TOI-4860 has a \textit{Gaia} DR3 RUWE value of 1.04 and therefore appears single. Additionally, TOI-4860 does not appear in the \textit{Gaia} DR3 non-single stars table \citep{gaiaDR3nonsingle}; as such, the \textit{Gaia} astrometric fit is fully consistent with a single-star model.

\subsubsection{High-resolution imaging} 
\begin{figure}[h]
    \centering
    \includegraphics[width=0.46\textwidth]{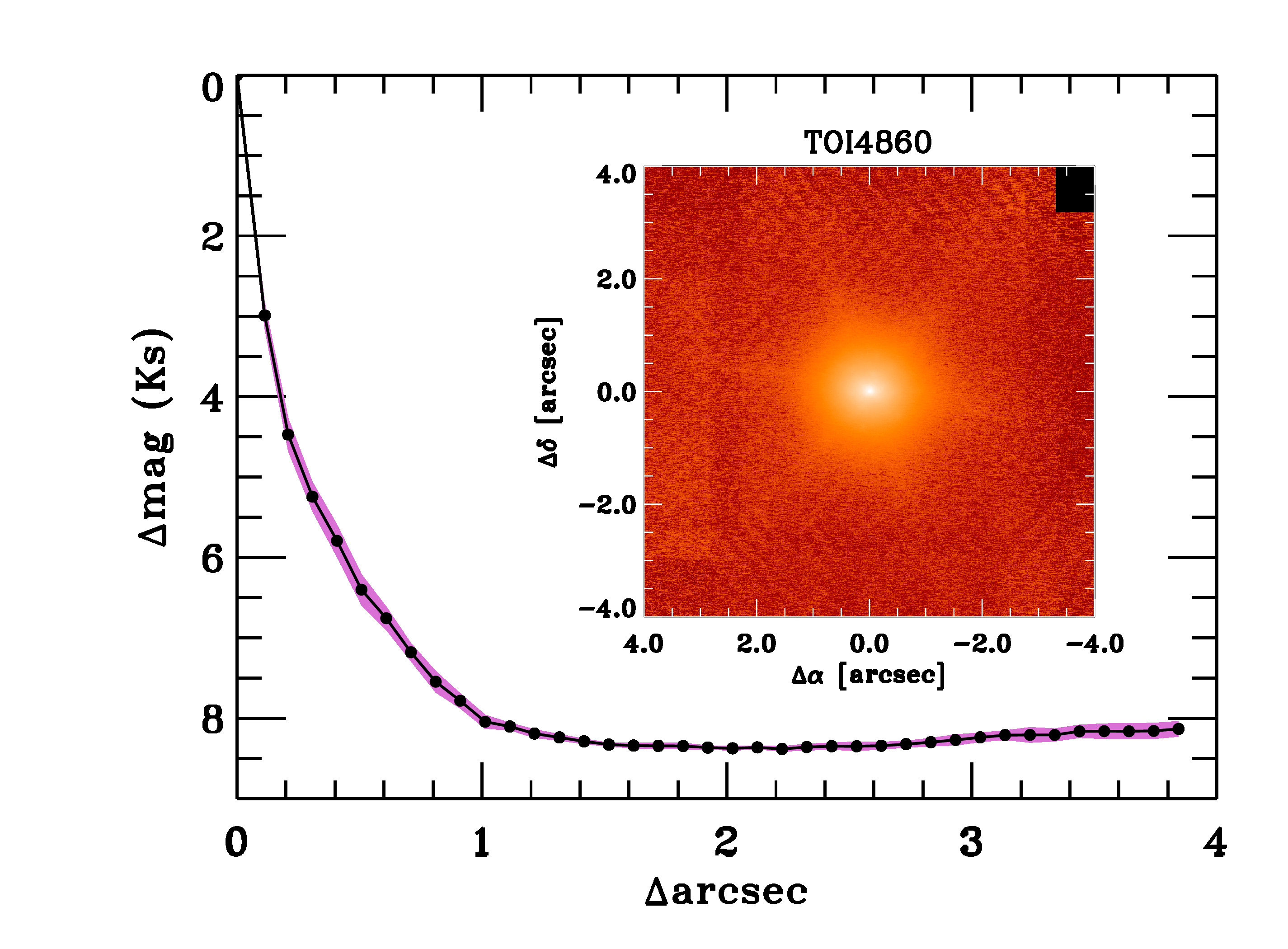}
    \caption{Companion sensitivity for the near-infrared AO imaging. The black points represent the 5$\sigma$ limits and are separated in steps of one full width at half maximum; the purple represents the azimuthal dispersion (1$\sigma$) of the contrast determinations (see the main text). The inset image is of the primary target and shows no additional close-in companions.}\label{fig:ao_contrast}  
        \vspace{-0.5em}
\end{figure}

As part of our standard process for validating transiting exoplanets to assess the possible contamination of bound or unbound companions on the derived planetary radii \citep{ciardi2015}, we also observed TOI-4860 with high-resolution near-infrared adaptive optics (AO) imaging at Keck Observatory. The observations were made with the Near Infra Red Camera 2 (NIRC2) instrument on Keck-II behind the natural guide star AO system \citep{wizinowich2000} on 10 June 2023 UT in the standard three-point dither pattern that is used with NIRC2 to avoid the left-lower quadrant of the detector, which is typically noisier than the other three quadrants. The dither pattern step size was $3\arcsec$ and was repeated twice, with each dither offset from the previous dither by $0.5\arcsec$. NIRC2 was used in the narrow-angle mode with a full field of view of $\sim10\arcsec$ and a pixel scale of approximately $0.0099442\arcsec$ per pixel. The Keck observations were made in both the $K_s$ filter $(\lambda_o = 2.146; \Delta\lambda = 0.311~\mu$m) and $J$ $(\lambda_o = 1.248; \Delta\lambda = 0.163~\mu$m) with an integration time in each filter of 20 and 40 seconds for a total of 180 and 360 seconds, respectively. Flat fields were generated from a median average of dark subtracted dome flats. Sky frames were generated from the median average of the nine dithered science frames; each science image was then sky-subtracted and flat-fielded. The reduced science frames were combined into a single combined image using a intra-pixel interpolation that conserves flux, shifts the individual dithered frames by the appropriate fractional pixels; the final resolution of the combined dithers was determined from the full-width half-maximum of the point spread function; 0.101\arcsec\ and 0.114\arcsec\ for the two filters, respectively. To within the limits of the AO observations, no stellar companions were detected. The final $5\sigma $ limit at each separation was determined from the average of all of the determined limits at that separation and the uncertainty on the limit was set by the root mean square (RMS) dispersion of the azimuthal slices at a given radial distance (Figure~\ref{fig:ao_contrast}).

\subsection{Ground-based photometry with ExTrA}
Exoplanets in Transits and their Atmospheres \citep[ExTrA;][]{Bon2015} is a near-infrared (0.85 to 1.55 $\mu$m) multi-object low-resolution spectrograph fed by three 60 cm telescopes located at La Silla Observatory in Chile. One partial and six full transits were observed using two or three of the ExTrA telescopes. The first of those transits happened to be partial due to the then uncertain ephemeris, and helped refine the ephemeris for the following observations.

We used 8$\arcsec$ aperture fibres and the lowest-resolution mode ($R$$\sim$20) of the spectrograph, a combination that is optimal for the target's magnitude, with an exposure time of 60~seconds. Five fibres are positioned in the focal plane of each telescope to select light from the target and four comparison stars. We chose comparison stars\footnote{2MASS J12145957-1303121, 2MASS J12154216-1243286, 2MASS J12130766-1243315, and 2MASS J12123143-1255153.} with 2-Micron All-Sky Survey (2MASS)~$J$ magnitudes \citep{Skrutskie2006} and effective temperatures \citep{gaia2018} similar to the target. The resulting ExTrA data were analysed using a custom data reduction software, described in more detail in \cite{Coi2021}.

\subsection{Radial velocities}
\subsubsection{SPIRou}

The SPectropolarim\`etre InfraROUge (SPIRou) instrument is a near-infrared high-resolution (980--2450~nm; $R=70,000$) high-stability ($\sim$1~\ms) velocimeter and spectropolarimeter installed at the Cassegrain focus of the 3.6 m Canada-France-Hawaii Telescope (CFHT) at Mauna Kea \citep{donati2020}. 
TOI-4860 was observed between March 2022 and July 2023 with SPIRou as part of the large programme SPIRou Legacy Survey (SLS; ID P42, PI: Jean-François Donati). It was observed at 28 observing epochs with four polarimetric exposures per epoch, except on 4 July 2023 when the sequence was interrupted after the first exposure. In total, 109 spectra were acquired with integration time changing from 184~s for the first 8 exposures to 602~s for the rest of the sequence. The median signal-to-noise ratio (S/N) per pixel measured in the middle of the $H$ band for the two exposure times were 9 and 28, respectively. 

To process the raw data obtained with SPIRou, we employed A PipelinE to Reduce Observations \citep[\texttt{APERO};][]{cook2022} version 0.7.275. The \texttt{APERO} software initially corrects detector effects, removes constant background thermal components, and detects bad pixels and cosmic-ray impacts. Subsequently, it calculates the position of echelle spectral orders and optimally extracts spectra from fibres A, B, A+B, and C into 2D order-separated e2ds\footnote{extracted 2-dimensional spectrum} and 1D order-merged s1d spectra. For the RV analysis, we use the A+B extraction (i.e. the combined extraction of the A and B science channels). A blaze function is derived from flat-field exposures. The wavelengths of the spectra are determined from a set of nightly calibrations following \cite{hobson2021}. An absolute calibration of wavelengths with respect to the Solar System barycentric rest frame is also performed using the current barycentric Earth radial velocity (BERV) and the barycentric Julian date (BJD) of each exposure with the \texttt{barycorrpy} code \citep{kanodia2018,wright2014}. Finally, \texttt{APERO} applies a three-step telluric correction that we summarise here with a complete description to come (Artigau et al.\ in prep.). First, the extracted spectrum is pre-cleaned by fitting a TAPAS atmospheric model \citep{bertaux2014}. Second, an empirical removal of residuals left after step one is performed using observations of fast-rotating hot stars observed with SPIRou at varying conditions (e.g. airmass and water column). Third, finite resolution artefacts from telluric modelling \citep{wang2022} are mitigated, leaving final residuals at the level of the \cite{artigau2014} principal component analysis method.

The line-by-line (LBL) method of \cite{Artigau2022} was then applied to the telluric-corrected spectra of TOI-4860 to measure its RVs. Several other TESS exoplanets were characterised with SPIRou using the LBL method in recent years: TOI-1759\,b \citep{martioli2022}, TOI-2136\,b, \citep{Gan2022}, TOI-1452\,b \citep{Cadieux2022}, and TOI-1695\,b \citep{Kiefer2023}. The LBL method is a statistical framework where the RV is obtained from the weighted average of thousands of independent velocity measurements on individual spectral lines over the full wavelength domain. Velocities are derived from the difference between the observations and a high-S/N template spectrum of the star (\citealt{Bouchy2001} formalism). In practice, one can use the template of any similar spectral type star in the LBL calculation. Since TOI-4860 is relatively faint, we chose to use instead the template of Gl 905, a bright M5V star observed in the SLS over three years. The final RVs are presented in Table~\ref{table.RVs} and have median uncertainties of respectively 34.3~\ms\ and 16.2~\ms\ for the 184~s and 602~s exposures.

The TOI-4860 SPIRou observations were performed in circular polarisation mode (Stokes~V), where each visit consists of a set of four exposures that provide a polarimetric spectrum. We used previously described methods \citep{Donati1997,martioli2020,martioli2022,cook2022}, implemented in the APERO pipeline, to calculate the polarimetric spectra of TOI-4860 and analyse the Stokes~V spectra using least squares deconvolution technique as in \cite{Donati1997}. With a typical S/N of only 30 to 40, and even using least squares deconvolution analysis, we obtained a mean polarimetric RMS per visit of $\sigma_{V}=0.2\pm0.1$\%. Such precision is sensitive to a disk-integrated longitudinal magnetic field on the stellar surface of the order of $B_\ell>20$~G. Non-detections suggest that TOI-4860 is not extremely active, but moderate magnetic activity cannot be completely ruled out.

\subsubsection{ESPRESSO}
We obtained seven spectra of TOI-4860 with the Echelle SPectrograph for Rocky Exoplanets and Stable Spectroscopic Observations \citep[ESPRESSO;][]{pepe2021} at the 8.2~m ESO Very Large Telescope (VLT) array, at the Paranal Observatory in Chile. The observations were obtained from January to February 2023 in Programme ID: 0110.C-4069 (PI: Jord\'an) as part of a long-running campaign to obtain mass measurements for giant planets transiting low-mass host stars. Initially, candidates for this campaign were drawn from the HAT-South survey \citep{bakos2013hatsouth}, and more recently from the TOI list \citep{guerrero2021} and our own dedicated search of the TESS FFI data \citep{bryant2023}. 

We observed TOI-4860 in the single Unit Telescope (UT), high-resolution (HR) mode (1$^{\prime\prime}$ fibre, R$\sim$140,000) over a spectral range from $\sim$380 to $\sim$780~nm. The spectrograph is contained in a temperature- and pressure-controlled vacuum vessel to avoid spectral drifts, and records cross-dispersed echelle spectra on two detectors. They were read out in the 2x1 binned mode, which presents a reduced level of continuum noise as compared to the unbinned readout mode. The calibration fibre (fibre B) was placed on sky in order to monitor and remove any possible contamination from the moon, as well as any sky emission lines.

The recorded spectra have a median S/N value of $\sim$\,8 at 550\,nm, with this value rising to $\sim$\,25 towards the red end of the spectrum. The spectra were reduced with the updated version of the dedicated Data Reduction Software (\drs), which, among other improvements, mitigates the bug in the exposure-meter data tables that affects the flux-weighted mid-exposure time calculation. The reduction includes all the standard steps, whereby the pipeline provides RV measurements by fitting a Gaussian function to the measured cross-correlation function \citep[CCF;][]{Baranne1996,Pepe2002}. This CCF is calculated using a stellar template matching closest the spectral type of the star (M4 in this case). The measured RVs are given in Table~\ref{table.RVs}.

\section{Stellar parameters}\label{section:stellarparameters}
TOI-4860 is an M-type star located at a distance of $80.42\pm 0.22$~pc \citep{gaia2016, gaiaDR3, lindegren2021}. The astrometry, photometry, and stellar parameters are reported in Table~\ref{tab:stellar_params}.\\

\begin{figure}[h]
   \centering
   \includegraphics[width=0.49\textwidth]{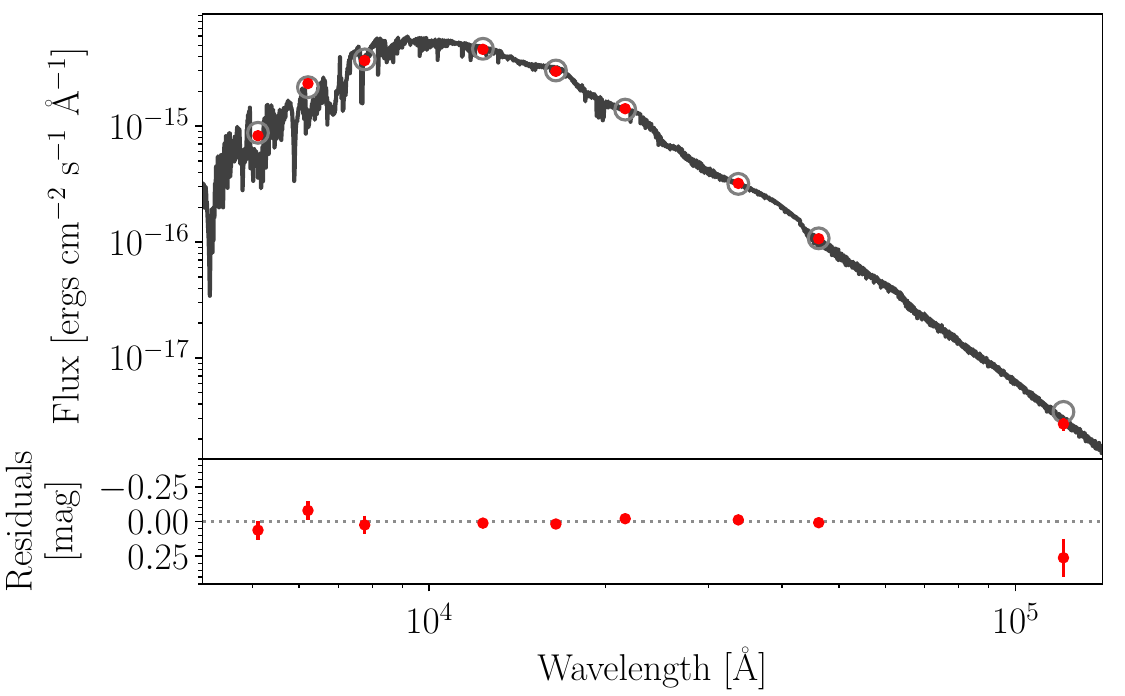}
      \caption{SED of TOI-4860. The solid line is the MAP PHOENIX/BT-Settl-interpolated synthetic spectrum, red circles are the absolute photometric observations (the jitter has been added quadratically to the data error bars), and grey open circles are the result of integrating the synthetic spectrum into the observed bandpasses.}
      \label{fig:sed}
\end{figure}

We derived the mass and radius of TOI-4860 from empirical relations based on luminosity. We used the \textit{Gaia}-corrected \citep{gaia2016,gaiaDR3,lindegren2021} parallax determination ($12.43\pm0.03$~mas) to compute the distance and an absolute magnitude of $M_{K_s}=6.648\pm0.027$. We then used the empirical relations of \citet{mann2019} and \citet{mann2015} with the metallicity dependence to derive a mass of $M_{\star}=0.340\pm0.009$~\Msun and a radius of $R_{\star}=0.358\pm0.015$~\Rsun, respectively. For the stellar radius, we used a systematic uncertainty floor of 4\% following \citet{tayar2022}. We derived an alternate stellar radius from the spectral energy distribution (SED) that we constructed using the magnitudes from \textit{Gaia} \citep{riello2021}, 2MASS \citep[][]{2mass,cutri2003}, and the Wide-field Infrared Survey Explorer \citep[WISE;][]{wise,cutri2013}. Those measurements are listed in Table~\ref{tab:stellar_params}. We modelled these magnitude measurements using the procedure described in \citet{diaz2014}, with the PHOENIX/BT-Settl \citep{allard2012} stellar atmosphere models. We used informative priors for the effective temperature ($T_{\mathrm{eff}}=3190\pm70$~K), and metallicity ($[\rm{Fe/H}]=0.27\pm0.12$~dex) derived from the co-added ESPRESSO spectra \citep[which we analysed with \specmatch;][]{yee2017}, and for the distance from \textit{Gaia}. We used uniform priors for the rest of the parameters. We used a jitter \citep{gregory2005} for each set of photometric bands (\textit{Gaia}, 2MASS, and WISE). The parameters, priors, and posteriors are listed in Table~\ref{table.sed}. The maximum a posteriori (MAP) model is shown in Fig~\ref{fig:sed}. The derived SED radius ($R_{\star}=0.371\pm0.010$~\Rsun) is compatible ($0.7\sigma$) with the radius computed above using an empirical radius-luminosity relation. The derived SED $T_{\mathrm{eff}}=3260\pm50$~K, which corresponds to an M3.5V spectral type \citep{pecaut2013}, is our adopted value.

\begin{table}[h]
  \setlength{\tabcolsep}{5.5pt}
      \caption[]{Stellar parameters for TOI-4860 (TIC 335590096, UCAC4 385-060574, 2MASS J12141555-1310290, WISE J121415.40-131029.3, \textit{Gaia} DR3 3571038605366263424).}
         \label{tab:stellar_params}
         \begin{tabular}{lcc}
            \hline
            \noalign{\smallskip}
            Parameter & \text{Value} & \text{Refs} \\
            \noalign{\smallskip}
            \hline
            \noalign{\smallskip}
            \textit{Astrometry} \\
            Right ascension (J2016), $\alpha$ & 12$^{\rm h}$14$^{\rm m}$15.34$^{\rm s}$ & 1 \\
            Declination (J2016), $\delta$ & $-$13$^{\rm o}$10'29.43'' & 1 \\
            Parallax, $\pi$ [mas] & $12.43\pm 0.03$ & 1,2 \\
            Distance, d [pc] & $80.42\pm 0.22$ & 1,2 \\
            Proper motion $\alpha$ [mas/year] & $-177.24\pm0.04$ & 1 \\
            Proper motion $\delta$ [mas/year] & $-5.43\pm0.03$ & 1 \\
            \noalign{\smallskip}
            \textit{Photometry} \\
            V [mag]& $16.47\pm0.02$ & 3 \\
            \textit{Gaia}-BP [mag] & $16.7328\pm0.0074$  & 1 \\
            \textit{Gaia}-G [mag]& $15.0774\pm0.0030$ & 1\\
            \textit{Gaia-R}P [mag] & $13.8435\pm0.0033$ & 1 \\
            TESS magnitude [mag]& $13.7726\pm0.0079$ & 3 \\
            J [mag]& $12.056\pm0.022$ & 4 \\
            H [mag]& $11.431\pm0.026$ & 4 \\
            $K_s$ [mag]& $11.175\pm0.026$ & 4 \\
            WISE-W1 [mag]  & $11.016\pm0.023$ & 5 \\
            WISE-W2 [mag]  & $10.886\pm0.020$ & 5 \\
            WISE-W3 [mag]  & $10.96\pm0.14$ & 5\\
            \noalign{\smallskip}
            \textit{Stellar parameters} \\
            Spectral type & M3.5 & 6 \\
            $M_{K_s}$ [mag]& $6.648\pm 0.027$ & 1,2,4,7 \\
            Stellar radius, $R_{\star}$ [\Rsun] & $0.358\pm0.015$ & 7,8 \\
            Stellar mass, $M_{\star}$ [\Msun] & $0.340\pm0.009$ & 7,9 \\
            Effective temperature, $T_{\rm eff}$ [K] & $3260 \pm 50$ & 7, SED \\
            Surface gravity, log g [cgs] &  $4.86 \pm 0.04$ & $R_\star$, $M_\star$ \\
            Metallicity, [Fe/H] [dex] & $0.27\pm 0.12$ & 7,10 \\            
            $\log R'_{HK}$ & $-5.65 \pm 0.12$ & 7 \\
            \noalign{\smallskip}
            \hline
        \end{tabular}
        \begin{tablenotes}
        \small
        \item References : 1) \cite{gaiaDR3}, 2) \cite{lindegren2021}, 3) \cite{stassun2019}, 4) \cite{cutri2003}, 5) \cite{cutri2013}, 6) \cite{pecaut2013}, 7) This work, 8) \cite{mann2015}, 9) \cite{mann2019}, 10) \cite{yee2017}.
        \end{tablenotes}
\end{table}

The flux around the calcium lines is low but using the co-added ESPRESSO spectra, we were able to measure the value of $\log R'_{HK}=-5.65\pm0.12$, derived from the calcium doublet. This translates to an estimated stellar rotation period of P$_{\rm rot}=114\pm22$~days using the activity-rotation relation of \citet{astudillo2017}.
Such a rotation period falls within the distribution of the `slow-rotator' population of mid-M-type dwarfs in the field \citep{Newton2017,Kiman2021,Popinchalk2021}; thus, the star is likely in the Skumanich-like spin-down phase, in which the braking law is strongly rotation-dependent and the period is a predictor of age. The effective temperature of TOI-4860 (3260~K) is at the poorly constrained cool end of the 4~Gyr gyrochronology calibration based on rotation periods of M67 M dwarf cluster members \citep{Dungee2022,Gaidos2023b}, but adopting this yields an age of $\sim$2.2~Gyr. The star could be older if it has gained angular momentum from the close-in planet via tides and is more rapidly rotating and active as a result \citep{ilic2022}.

\section{Analysis and modelling}\label{section.analysis}
We used the software package \juliet \citep{espinoza2019} to model the photometric and RV data. The algorithm is built on many publicly available tools for the modelling of transits \citep[\batman,][]{kreidberg2015}, RVs \citep[\radvel,][]{fulton2018}, and Gaussian processes (GPs) (\george, \citealt{ambikasaran2015}; \celerite, \citealt{foreman2017}). In order to compare different models, \juliet efficiently computes the Bayesian evidence ($\mathcal{Z}$) using \dynesty \citep{speagle2020}, a python package to estimate Bayesian posteriors and evidence using nested sampling methods. Instead of starting with an initial parameter vector centred on a likelihood maximum discovered through optimisation techniques, nested sampling algorithms sample directly from the given priors.

\subsection{Transit chromaticity with TESS and ExTrA photometry}

\begin{figure}[h]
   \centering
   \includegraphics[width=0.48\textwidth]{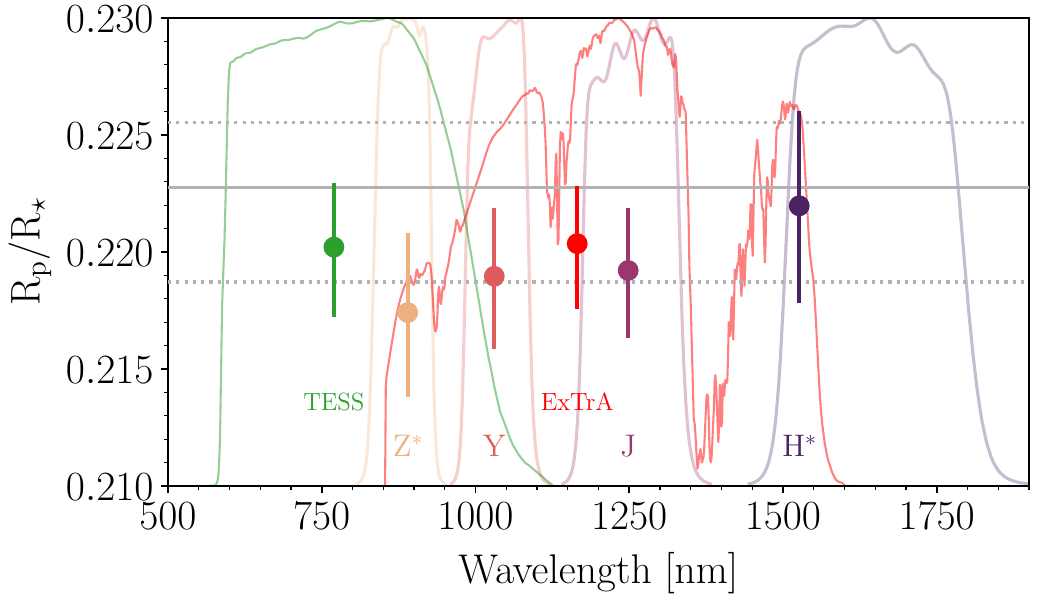}
      \caption{Posterior distribution comparison for $R_p/R_\star$ (error bars) computed for different bands (shown and labelled with different colours). TESS, ExTrA (0.85 to 1.55~$\mu$m), Z$^{*}$, Y, J, and H$^{*}$ data are fitted jointly with a different $R_p/R_\star$ parameter for each of the datasets. The horizontal solid and dotted grey lines represent the posterior median and 68.3\% CI of $R_p/R_\star$ inferred in Sect.~\ref{sect:join}.}
      \label{fig:RpRsRatio}
\end{figure}

A wavelength-dependent transit depth would suggest a false positive scenario involving an eclipsing binary. We therefore performed a joint fit of both TESS and ExTrA photometry. To push further the chromatic analysis, in addition to ExTrA's full wavelength range (0.85 to 1.55~$\mu$m) light curve, we synthesised from the ExTrA spectrophotometry photometry in the UKIRT-WFCAM filters,\footnote{Retrieved from the SVO Filter Profile Service \citep[\url{http://svo2.cab.inta-csic.es/theory/fps/},][]{rodrigo2012,rodrigo2020}.} to produce four additional light curves: a truncated Z band (Z$^{*}$), a $Y$, a $J$, and a truncated H band (H$^{*}$), shown in Fig.~\ref{fig:ExTrA_ZYJH}. The analysis is similar to that presented in Sect.~{\ref{sect:join}} except that RVs are not used and the eccentricity is fixed to zero. All planetary parameters except for the planet-to-star radius ratio ($R_p/R_\star$) are common to all datasets, to allow for a different transit depth at each band, whose posteriors' median and 68.3\% credible interval (CI) are listed in Table~\ref{table.RpRsRatio}. As shown in Fig.~\ref{fig:RpRsRatio} the photometry for all of these bandpasses is consistently modelled with a similar $R_p/R_\star$. No chromaticity is seen, lending confidence in the planet detection rather than a scenario with a blended binary\footnote{An equivalent analysis was done with just the first full ExTrA transit and the first two TESS sectors prior to the acquisition of RV observations.}.

\begin{table}
  \setlength{\tabcolsep}{5pt}
\renewcommand{\arraystretch}{1.25}
\centering
\caption{Inferred $R_p/R_\star$ in different bands.}\label{table.RpRsRatio}
\begin{tabular}{lrcc}
\hline
\hline
Band & $\lambda_{\rm pivot}$~[nm] & Prior &  Median and 68.3\% CI  \\
\hline
TESS           &   769.8  & $U(0.1, 0.3)$ & $0.2202^{+0.0027}_{-0.0030}$ \\
ExTrA          &  1165.4  & $U(0.1, 0.3)$ & $0.2204^{+0.0025}_{-0.0028}$ \\
ExTrA $Z^{*}$  &   889.8  & $U(0.1, 0.3)$ & $0.2174^{+0.0034}_{-0.0036}$ \\
ExTrA $Y$        &  1030.5  & $U(0.1, 0.3)$ & $0.2190^{+0.0029}_{-0.0031}$ \\
ExTrA $J$        &  1248.3  & $U(0.1, 0.3)$ & $0.2192^{+0.0027}_{-0.0029}$ \\
ExTrA $H{*}$  &  1526.4  & $U(0.1, 0.3)$ & $0.2220^{+0.0041}_{-0.0042}$ \\
\hline
\end{tabular}
\tablefoot{The table lists: band, $\lambda_{\rm pivot}$ \citep{koornneef1986}, prior, posterior median, and 68.3\% CI. $U(a, b)$: A uniform distribution defined between a lower $a$ and an upper $b$ limit.}
\end{table}

\subsection{Transit timing variations}

Transit timing variations (TTVs) would indicate the presence of additional planets \citep{agol2005,holman2005} or orbital decay \citep{patra2017}. We analysed the TESS and ExTrA transit photometry with two models: one with a linear ephemeris and a second that allows for the time of individual transits to vary (with uniform priors of $\pm$10~minutes width around the linear ephemeris of the first model). The resulting TTVs are shown in Fig.~\ref{fig:TTV}. A model comparison strongly favours \citep{kass1995} the model with a linear ephemeris (P, T$_0$) over the one with TTVs (log-Bayes factor $\mathcal{Z}_{\rm P,T_0}/\mathcal{Z}_{\rm TTVs}$ of $92.5\pm0.8$).

\subsection{Radial velocity}\label{section.RV}

The RVs are listed in Table~\ref{table.RVs}. One ESPRESSO and six SPIRou observations were taken during transit. The $\log R'_{HK}$ is indicative of a slow rotator, which would imply a Rossiter–McLaughlin effect \citep{rossiter1924,mcLaughlin1924} semi-amplitude of just $\sim$5~\ms\ (based on P$_{\rm rot}$, estimated from the $\log R'_{HK}$, we computed a \vsini\ of $\sim$160~\ms, assuming $i_\star = 90$\degree). Except for in regards to the discussion on a potential system misalignment (Sect.~\ref{section.results}, Fig.~\ref{fig:RM}), we excluded the ESPRESSO measurement taken during the transit from our analyses. However, we included the SPIRou measurements taken during the transit in all our analyses, as the expected amplitude of the Rossiter-McLaughlin effect is smaller than their accuracy. For the SPIRou data, we used the nightly weighted average (we excluded the spectra on BJD 2460096.806090 because it is an outlier, and BJD 2460064.896367 and 2460064.903591 because of a low S/N of about 2.5 in the middle of the $H$ band).
We first analysed the RV data alone to choose the model for the RVs in the global fit.

Figure~\ref{fig:SPIRou_periodogram} shows the periodogram of SPIRou RVs, whose highest peak corresponds to the period of TOI-4860\,b. The residuals, after the signal from TOI-4860\,b has been removed, exhibit additional signals over a longer timescale. This finding is also supported by the ESPRESSO data, for which a model with a circular planet (with priors on ephemeris from the transit observations) and a drift in RV are strongly favoured relative to the same model without drift (log-Bayes factor of $3.9\pm0.3$). We conducted a joint modelling of the SPIRou and ESPRESSO data to search for a second planet. We assumed a circular orbit for the transiting planet, and added a Keplerian with a period range between 10 and 1000 days. The marginal posterior distribution of the trial period shows a peak at 427~days (see Fig.~\ref{fig:SPIRou_periodogram}). The log-Bayes factor for this model with respect to the one-planet model is $12.3\pm0.4$, which can be interpreted as very strong evidence in favour of a second planet in the system (planet c candidate). We therefore used a model with two Keplerians for the joint fit of transit photometry and RVs. To confirm whether the additional signal in the RVs corresponds to a second planet in the system, more data are needed. The current time-span of 480 days is too close to the proposed period, and there is poor phase coverage.

The posterior of the SPIRou jitter in the two-planet modelling is $27 \pm 6~\ms$, doubling the SPIRou uncertainties. This jitter is probably inherent to the SPIRou data since the ESPRESSO data do not support it.

\subsection{Joint fit of all data}\label{sect:join}

\begin{figure*}[h]
   \centering
   \includegraphics[width=1.0\textwidth]{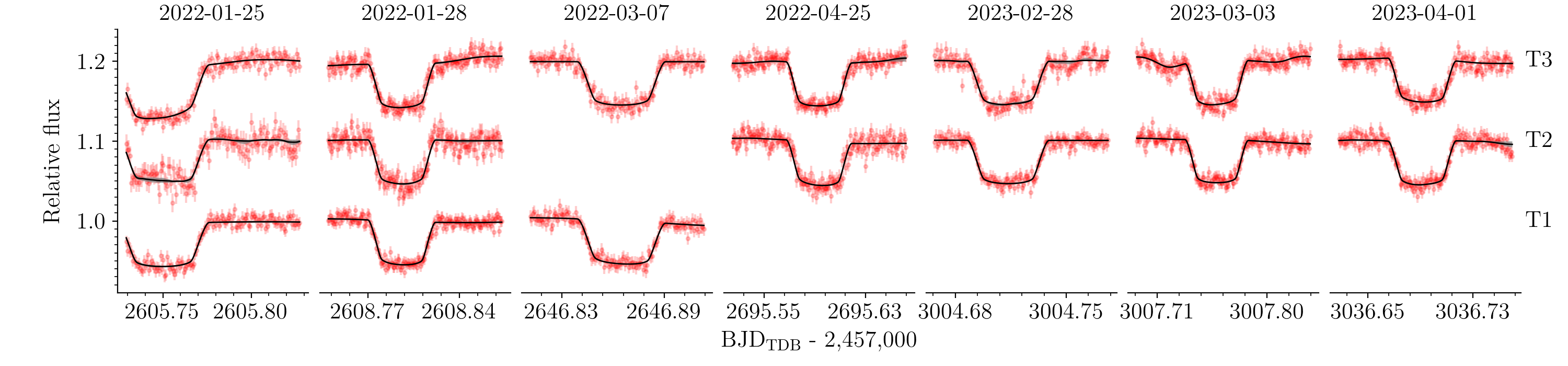}
   \includegraphics[width=1.0\textwidth]{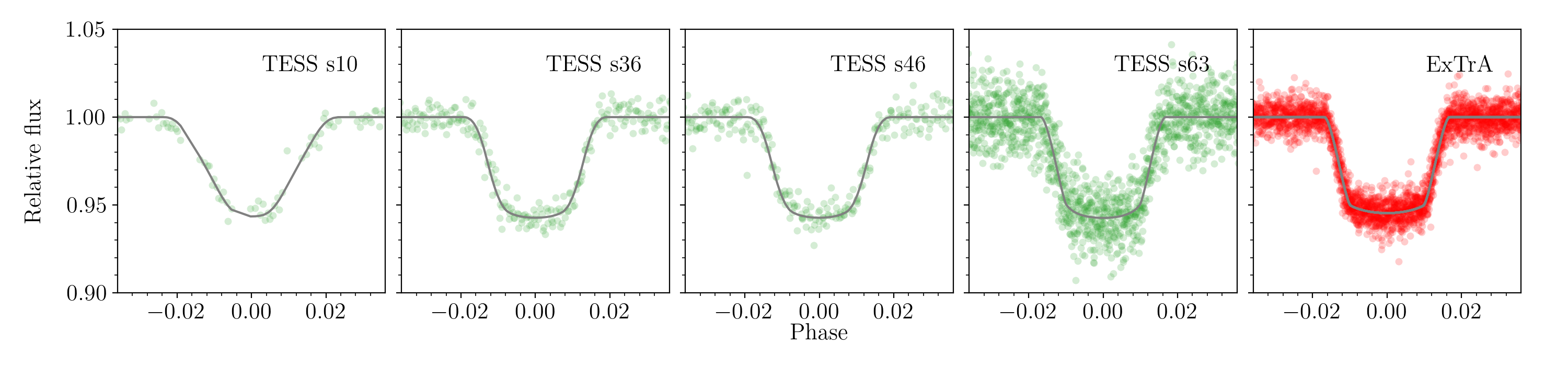}
   \includegraphics[width=1.0\textwidth]{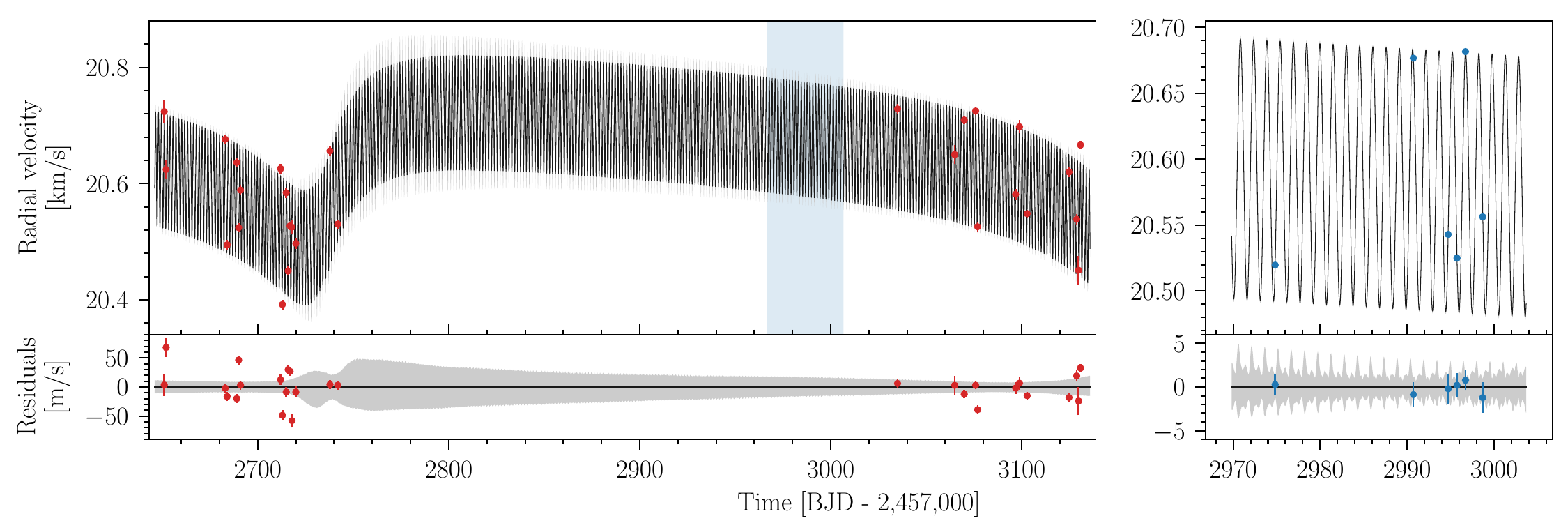}
      \caption{Modelling of transit photometry and RVs of TOI-4860. {\it Top panel:} ExTrA transit photometry observations of TOI-4860\,b. Each column corresponds to a night (labelled with format YYYY-MM-DD) and each line to an ExTrA telescope (labelled T1, T2, and T3) that is offset vertically for clarity. For each transit, the median model (black line) and the 68\% CI (grey band, barely visible) computed from 1000 random posterior samples are shown. {\it Middle panel:} TESS (separated by sector) and ExTrA transits detrended with the MAP model (whose transit component is shown as a grey line) and phase-folded. TESS sector 10 looks more V-shaped due to the 30-minute sampling. {\it Bottom-left panel}: SPIRou RVs (red error bars), median model (black line), and 68\% CI (grey band). Residuals from the median model are shown. {\it Bottom-right panel}: Same as the bottom-left panel but for ESPRESSO RVs (blue error bars). The time span of the ESPRESSO panel is highlighted as a blue band in the SPIRou panel.}
      \label{fig:global}
\end{figure*}

\begin{figure}
  \centering
  \includegraphics[width=0.47\textwidth]{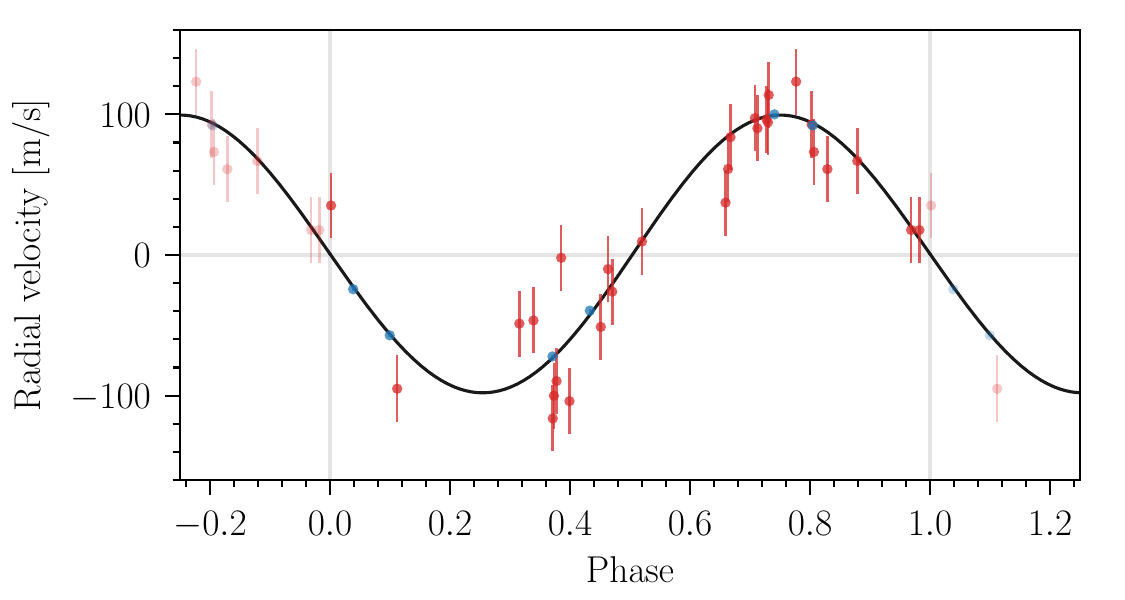}
  \includegraphics[width=0.47\textwidth]{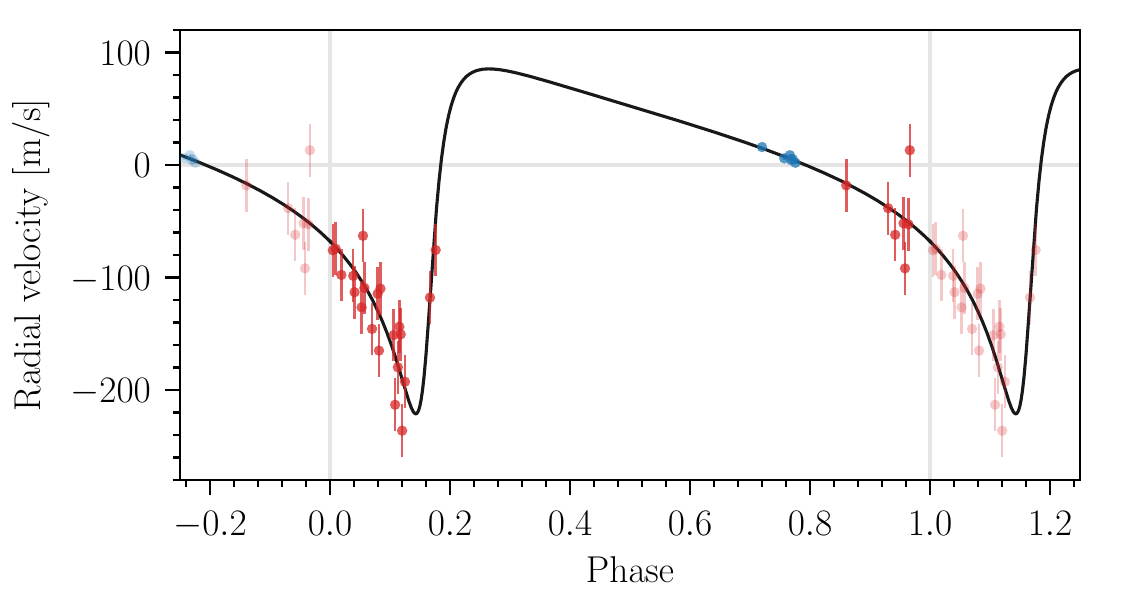}
  \caption{Phased RVs. {\it Top panel:} SPIRou (red error bars) and ESPRESSO (blue error bars) RVs corrected for the MAP Keplerian orbit of planet~c and phase-folded to the period of planet~b (the MAP jitter has been added quadratically to the data error bars). The black line is the best-fit Keplerian model from the joint fit. {\it Bottom panel:} Idem for the planet c candidate.} 
  \label{fig:RV_phase}
\end{figure}

We conducted an analysis of the transit photometry from TESS (data were extracted around three transit durations at the centre of each transit) and ExTrA (spanning the full wavelength range from 0.85 to 1.55~$\mu$m), as well as the RVs from SPIRou and ESPRESSO. A GP with an approximate Matern kernel was utilised to model the residuals in the transit photometry. Each transit observation had distinct kernel hyperparameters, except for the TESS data, where common kernel hyperparameters were used within each sector. For TESS sectors 10, 36, and 46, we over-sampled the model in time and adjusted for the observation's integration time through binning \citep{kipping2010}. The RV model comprised two Keplerians. We sampled from the posterior using \dynesty \citep{speagle2020}. Table~\ref{table.params} presents the prior, median, and 68\% CI of the marginal distributions of the inferred system parameters. The stellar mass and radius, as determined in Sect.~\ref{section:stellarparameters}, were used to calculate the planetary mass and radius of TOI-4860\,b. Figures~\ref{fig:global} and \ref{fig:RV_phase} display the datasets and the model derived from this analysis. The posterior of the stellar density ($11.7 \pm 0.8~\mathrm{g\;cm^{-3}}$) is consistent with the value derived from the stellar mass and radius in Sect.~\ref{section:stellarparameters} ($10.4 \pm 1.5~\mathrm{g\;cm^{-3}}$).

We examined the transit windows of the candidate planet~c over the span of available observations, and all of them fall at least 2$\sigma$ away from a TESS sector. Additionally, we used \rebound \citep{rein2012} to estimate the TTVs of TOI-4860\,b, due to the presence of the candidate planet~c, for the median values of the posterior. The semi-amplitude of the TTVs is 1.6~seconds with a period corresponding to the candidate planet~c, and is dominated by the light-time effect \citep{irwin1952}. This signal is well below the precision (18 or 56~seconds at best for a transit observed with ExTrA or TESS, respectively) and dispersion of the measured TTVs (Fig.~\ref{fig:TTV}).


\begin{table*}
  \small
  \setlength{\tabcolsep}{5pt}
\renewcommand{\arraystretch}{1.05}
\centering
\caption{Inferred system parameters.}\label{table.params}
\begin{tabular}{lccc}
\hline
\hline
Parameter & Units & Prior &  Median and 68.3\% CI  \\
\hline
\emph{\bf Star} \\
Mean density, $\rho_{\star}$     & [$\mathrm{g\;cm^{-3}}$] & $U(5, 20)$ & $11.7 \pm 0.8$ \\
$q_1$ TESS &                     & $U(0, 1)$ & $0.34^{+0.26}_{-0.16}$                      \\
$q_2$ TESS &                     & $U(0, 1)$ & $0.29^{+0.3}_{-0.17}$                     \\
$q_1$ ExTrA  &                   & $U(0, 1)$ & $0.136^{+0.21}_{-0.09}$                     \\
$q_2$ ExTrA  &                   & $U(0, 1)$ & $0.26^{+0.4}_{-0.19}$                              \smallskip \\

\emph{\bf Planet\,b} \\
Semi-major axis, $a$                   & [au]             & & $0.01808 \pm 0.00015$ \\
Eccentricity, $e$                      &                  & & < 0.035$^{(a)}$, ($0.008^{+0.013}_{-0.006}$)  \\
Argument of pericentre, $\omega$       & [\degree]        & & $240 \pm 130$ \\
Inclination, $i_p$                     & [\degree]        & & $88.5^{+0.6}_{-0.4}$ \\
Radius ratio, $R_{\mathrm{p}}/R_\star$ &                  & & $0.2228^{+0.0028}_{-0.004}$ \\
Scaled semi-major axis, $a/R_{\star}$  &                  & & $11.0 \pm 0.4$ \\
Impact parameter, $b$                  &                  & & $0.29^{+0.06}_{-0.11}$\\ 
Transit duration, $T_{14}$             & [h]              & & $1.231 \pm 0.011$ \\
T$_0$ \;-\;2\;460\;000 & [BJD$_{\mathrm{TDB}}$]   & $U(36.6907, 36.6918)$ & $36.69124 \pm 0.00010$ \\
Orbital period, $P$                    & [d]     & $U(1.522757, 1.522761)$     & $1.5227591 \pm 3\times 10^{-7}$ \\
RV semi-amplitude, K                   &[\ms]     & $U$(90, 110)   & $98.9 \pm 1.3$ \\
Radius, $R_{\mathrm{p}}$               &[\Renom]           & & $8.7 \pm 0.4$ \\
                                       &[\RJnom]      & & $0.77 \pm 0.03$ \\
Mass, $M_{\mathrm{p}}$                 &[\MEarth]          & & $86.7 \pm 1.9$ \\
                                       &[\Mjup]      & & $0.273 \pm 0.006$ \\
Mean density, $\rho_{\mathrm{p}}$ &[$\mathrm{g\;cm^{-3}}$]& & $0.75 \pm 0.09$ \\
Surface gravity, $\log$\,$g_{\mathrm{p}}$ &[cgs]          & & $3.06 \pm 0.03$ \\
Equilibrium temperature, T$_{\rm eq}$  & [K]              & & $694 \pm 15$ \\
Insolation flux                        & [F$_{\rm E}$]    & & $42.5 \pm 1.9$ \\
$r_1$                              &    & $U(0, 1)$ & $ 0.53^{+0.04}_{-0.07}$ \\
$r_2$                              &    & $U(0.20, 0.24)$ & $0.2228^{+0.0028}_{-0.004}$ \\
$\sqrt{e}\cos{\omega}$             &    & $U(-1, 1)$ & $-0.02 \pm 0.06$ \\
$\sqrt{e}\sin{\omega}$             &    & $U(-1, 1)$ & $-0.02 \pm 0.10$ \smallskip \\

\emph{\bf Candidate planet\,c} \\
Semi-major axis, $a$                   & [au]             & & $0.776 \pm 0.011$ \\
Eccentricity, $e$                      &                  & & $0.66 \pm 0.09$  \\
Argument of pericentre, $\omega$       & [\degree]        & & $222 \pm 14$ \\
Scaled semi-major axis, $a/R_{\star}$  &                  & & $466 \pm 22$ \\
T$_0$ \;-\;2\;460\;000 & [BJD$_{\mathrm{TDB}}$]   & $U(-200, 350)$ & $92 \pm 19$ \\
Orbital period, $P$                      & [d]              & $U(350, 550)$     & $427 \pm 7$ \\
RV semi-amplitude, K             &[\ms]     & $U$(0, 250)                        & $121 \pm 27$ \\
Minimum mass, $M_{\mathrm{p}}\sin{i}$                 &[\Mjup]      & & $1.66 \pm 0.26$ \\
Equilibrium temperature, T$_{\rm eq}$  & [K]              & & $107 \pm 3$ \\
Insolation flux                        & [F$_{\rm E}$]    & & $0.0230 \pm 0.0012$ \\
$\sqrt{e}\cos{\omega}$             &    & $U(-1, 1)$ & $-0.59^{+0.14}_{-0.09}$ \\
$\sqrt{e}\sin{\omega}$             &    & $U(-1, 1)$ & $-0.54 \pm 0.18$ \smallskip \\

\emph{\bf RV} \\
SPIRou jitter      & [\ms]       &  $J(0.01, 80)$ & $27 \pm 5$ \\
SPIRou offset    & [\ms]     & $U(20500, 20800)$ & $20662 \pm 16$ \\
ESPRESSO jitter      & [\ms]       &  $J(0.01, 10)$ & $0.14^{+0.7}_{-0.12}$ \\
ESPRESSO offset & [\ms]     & $U(20400, 20650)$ & $20571 \pm 9$ \smallskip\\

\emph{\bf Photometry} \\
Offset relative flux & [Relative flux]  & $N(0, 0.01)$ & $^{(b)}$ \\
Jitter      & [ppm]            & $J(1, 1000)$ & $^{(b)}$ \\
Amplitude of the GP  & [Relative flux]  & $J(10^{-6}, 0.1)$ & $^{(b)}$ \\
Timescale of the GP  & [days]           & $J(0.001, 10)$ & $^{(b)}$ \\
\hline
\end{tabular}
\tablefoot{The table lists: prior, posterior median, and 68.3\% CI. Parameters without prior are derived parameters. $^{(a)}$~Upper limit, 95\% confidence. $^{(b)}$~The parameters listed for the photometry are different for each TESS sector and for each individual transit observed with each of the telescopes of ExTrA. The parameters $q_1$ and $q_2$ are the quadratic limb-darkening coefficients parameterised using \citet{kipping2013}. The parameters $r_1$ and $r_2$ are the impact parameter and transit depth parameterised using \citet{espinoza2018}. The planetary equilibrium temperature is computed for zero albedo and full day/night heat redistribution. IAU 2012: \rm{au} = 149$\;$597$\;$870$\;$700~\rm{m}$\;$. IAU 2015: \Rnom = 6.957\ten[8]~\rm{m}, \Renom~=~6.378$\;$1\ten[6]~\rm{m}, \GMnom = 1.327$\;$124$\;$4\ten[20]~$\rm{m^3\;s^{-2}}$, \GMenom = 3.986$\;$004\ten[14]~$\rm{m^3\;s^{-2}}$, $\RJnom$ = 7.149$\;$2\ten[7]~\rm{m}, \GMJnom = 1.266$\;$865$\;$3\ten[17]~$\rm{m^3\;s^{-2}}$. \Msun = \GMnom/$\mathcal G$, \MEarth = \GMenom/$\mathcal G$, \Mjup = \GMJnom/$\mathcal G$. CODATA 2018: $\mathcal G$ = 6.674$\;$30\ten[-11]~$\rm{m^3\;kg^{-1}\;s^{-2}}$. $U(a, b)$: A uniform distribution defined between a lower $a$ and an upper $b$ limit. $J(a, b)$: Jeffreys (or log-uniform) distribution defined between a lower $a$ and upper $b$ limit. $N(\mu, \sigma)$: Normal distribution prior with mean $\mu$, and standard deviation $\sigma$.}
\end{table*}

\section{Results and discussion}\label{section.results}
In this paper we present the discovery and characterisation of a giant planet, TOI-4860\,b, transiting an M3.5 dwarf with a period of 1.52~days. The exoplanet was detected by the TESS mission, then characterised from ground-based photometric follow-up using ExTrA and from precise RV measurements by both SPIRou and ESPRESSO.

\begin{figure}
  \centering \includegraphics[width=0.48\textwidth]{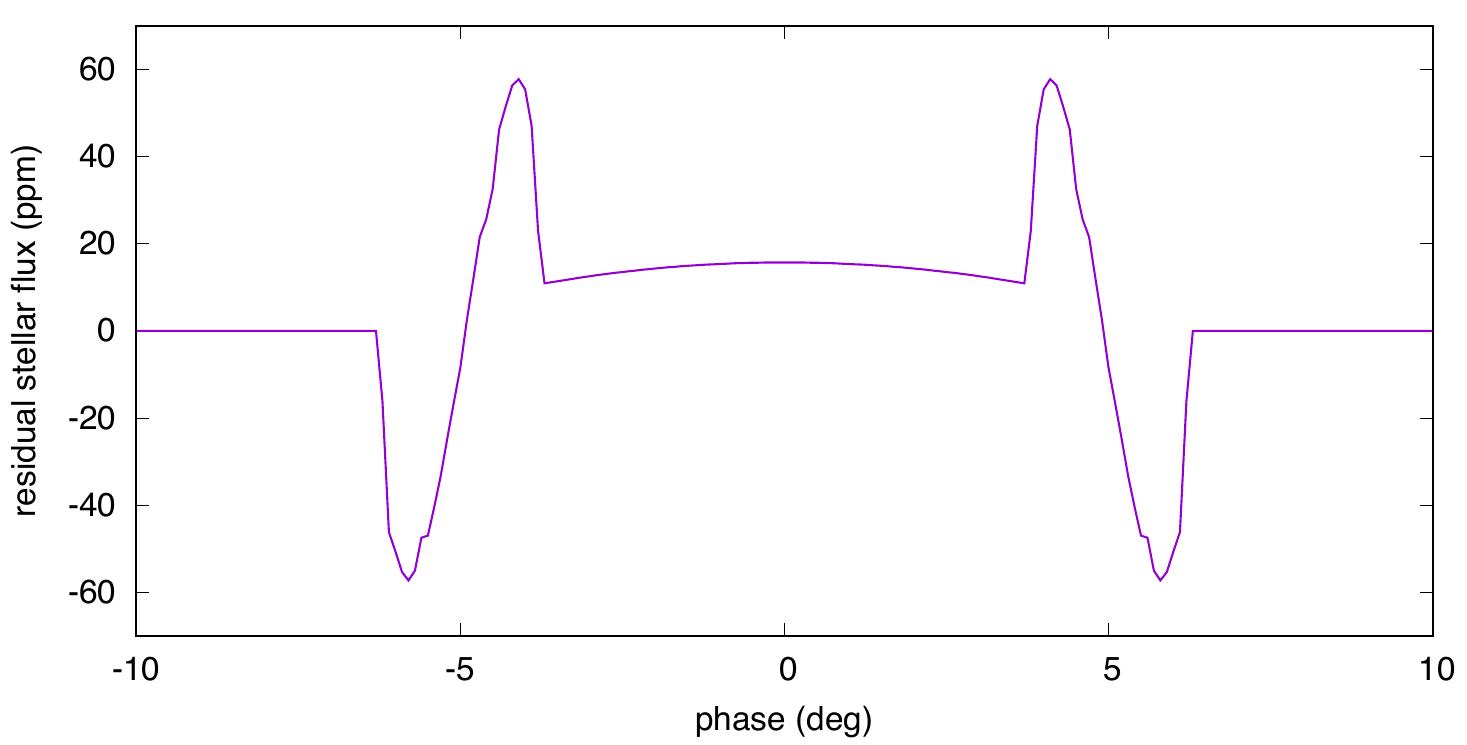}
  \caption{Difference between the transit light curves of an ellipsoidal and a spherical planet for TOI-4860\,b, obtained with uniform stellar flux.} 
  \label{fig:deformation}
\end{figure}

The giant planet is close to its host star, at only 1.8 times the Roche limit \citep[see][]{Chandrasekhar_1987}. Therefore, we expect that the system undergoes strong tidal interactions, which distort the planet and lead to orbital decay.
Adopting a radial fluid Love number identical to that of Jupiter, $h_f = 1.5$, we estimate a difference between the longest and the shortest radius of the planet of about 3\% \citep{Correia_2014}. In addition, since the planet's mean radius is about 22\% of the stellar radius, we expect $\sim60$~ppm discrepancies with respect to a spherical planet during transit ingress and egress (see Fig.~\ref{fig:deformation}). These variations are similar to those observed for WASP-103\,b \citep{Akinsanmi_etal_2019, Barros_etal_2022} and can thus be used to probe the interior structure of the planet using the transit light curves \citep{Ragozzine_Wolf_2009, Correia_2014}.
The tidal dissipation of M-dwarf stars is unknown, but often believed to be similar to that of other convective stars \citep[e.g.][]{Mathis_2015}. Assuming a tidal quality factor of $Q_*'=10^5$, we estimate a variation in the orbital period of about $\dot P \sim 10^{-9}$~day/yr \citep{Maciejewski_etal_2018}. This rate is not detectable with current instrumentation, but a direct estimation of $\dot P$ can still be used to put a lower limit on the tidal $Q$ factor of the star \citep[e.g.][]{Barros_etal_2022}.

\begin{figure*}
    \centering
    \includegraphics[width=0.66\columnwidth]{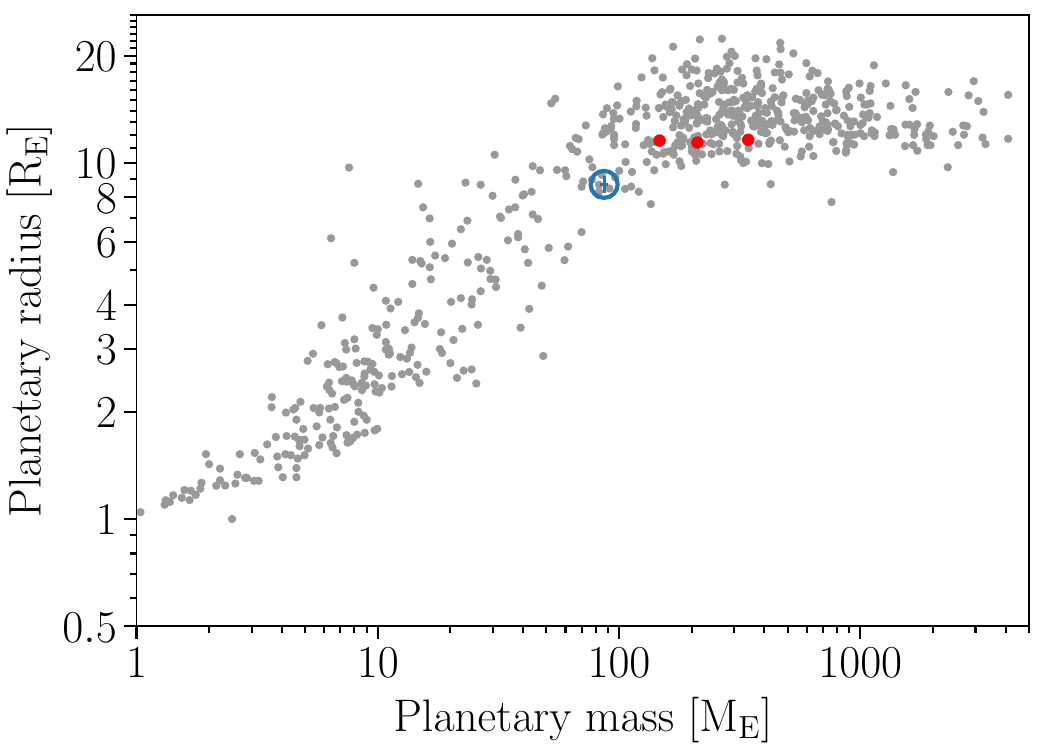}
    \includegraphics[width=0.66\columnwidth]{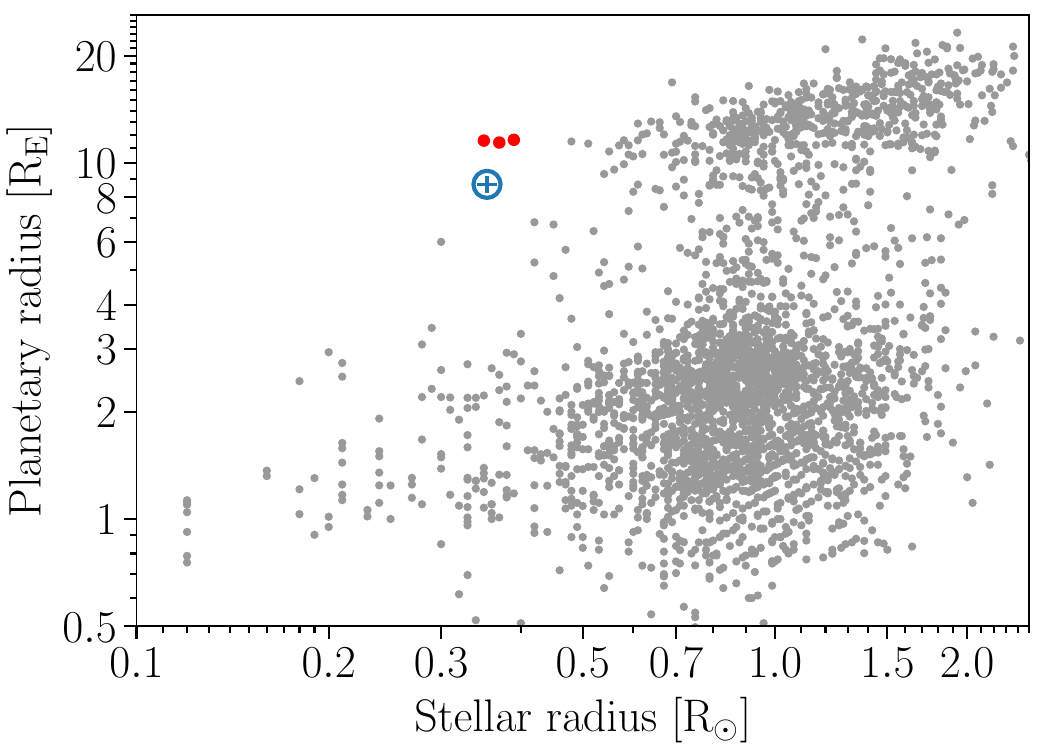}
    \includegraphics[width=0.66\columnwidth]{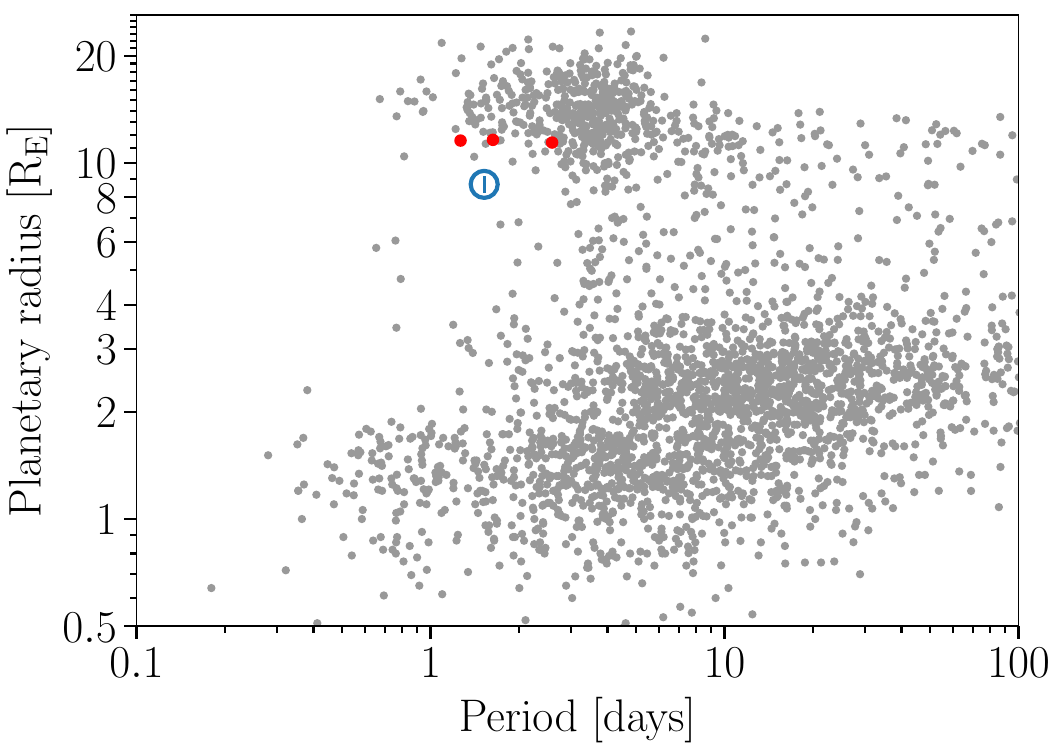}
    \caption{Comparison of TOI-4860\,b with known exoplanets. {\it Left}: Mass--radius diagram of known exoplanets. Grey dots are transiting planets listed in the NASA Exoplanet Archive (\url{https://exoplanetarchive.ipac.caltech.edu/}) with planetary radius and mass uncertainties below 20\%. TOI-519, TOI-3235, and TOI-5205 are shown (from left to right) as red dots. The blue open circle with error bars marks the position of TOI-4860\,b. {\it Centre}: Idem for planetary versus stellar radii (restricted to planets with planetary radius uncertainties below 20\%). {\it Right}: Idem for planetary radii versus orbital period.}
    \label{fig:MR_RpRs}
\end{figure*}

A long-standing prediction from core accretion planet formation is that giant planets do not form around low-mass stars \citep{laughlin2004}. In particular, the models of \citet{burn2021} are unable to form giant planets around stars with masses $M_* \lesssim 0.4 M_\odot$. The lack of such systems from the first ground-based transit surveys, such as WASP \citep{wasp,gaidos2014} and HATNet \citep{hatnet}, seemed to confirm this prediction, although the magnitude range of these surveys meant very few low-mass stars were monitored to the precision needed to find transiting exoplanets. 
A new generation of ground-based transit surveys, with increased telescope apertures \citep[e.g. HAT-South and NGTS;][]{bakos2013hatsouth, ngts1}, began to find transiting gas giant planets around lower-mass stars \citep{hats6,ngts1,hats71,jordan2022hats74_77}
but still not below the mass where core-accretion models have trouble forming giants. The TESS mission, with its near all-sky coverage and red wavelength sensitivity, is very well placed to detect giant planets transiting low-mass stars. This has allowed the discovery of transiting giants hosted by stars with masses $M_\star\lesssim0.4$~\Msun (TOI-5205, \citealt{kanodia2023}; TOI-3235, \citealt{hobson2023}; and TOI-519, \citealt{parviainen2021,kagetani2023}).
The TOI-4860 system, an M3.5V star that hosts a giant planet with a radius of $0.77\pm0.03$~\Rjup\ and a mass of $0.273\pm0.006$~\Mjup\ (see Fig.~\ref{fig:MR_RpRs}) at the border of the hot Neptune desert \citep[e.g.][]{lecavelier2007}, joins this short list of discoveries. With a mass of $0.340\pm0.009$~\Msun and a radius of $0.358\pm0.015$~\Rsun, its host star is well within the mass range that challenges core accretion.

These recent discoveries have allowed for the occurrence rates of giant planets around low-mass stars to be calculated, at least for short orbital period systems. The results from \citet{gan2023} and \citet{bryant2023} suggest that short-period giant planets occur around early-type M dwarfs at approximately half the frequency they do around solar-mass stars, and about a quarter the frequency when comparing mid to late M-dwarf hosts (0.088$-$0.4~\Msun), such as TOI-4860, to solar-mass stars. These results will become more concrete as more systems similar to TOI-4680 are confirmed.

TOI-4860 appears to be metal-rich ($\rm [Fe/H] = 0.27\pm0.12$), like other M dwarfs that host giant planets \citep[see Fig.~3 of][]{kagetani2023}. This suggests, as for FGK stars \citep{fischer2005}, a correlation between the metallicity of a star and the likelihood of its hosting giant planets. 

TOI-4860\,b has a high transmission spectroscopy metric \citep{kempton2018} of $183\pm15$, among the highest for its planet size. Moreover, based on the host star's $\log R'_{HK}$ ($-5.65 \pm 0.12$), the lack of flares in the TESS light curve, and the small ESPRESSO RV jitter (< 1.9~\ms\ at 95\% CI), we conclude that TOI-4860 currently exhibits low levels of activity. The planet's atmospheric composition can reveal insights into its formation and evolution history and is a key constraint for models of the planetary interior. 
Another constraint that can be placed upon the formation history could be the spin-orbit angle. The expected semi-amplitude of the Rossiter–McLaughlin effect \citep{rossiter1924,mcLaughlin1924} for TOI-4860\,b is $\sim$5~\ms, well above ESPRESSO capabilities \citep[e.g.][]{bourrier2022}. The one ESPRESSO observation obtained during transit does not seem compatible with an aligned orbit (Fig.~\ref{fig:RM}), but this will have to be confirmed with further observations.

TOI-4860 is located precisely at the fully convective boundary \citep{chabrier1997}, which can influence the star’s angular momentum evolution \citep{irwin2011}. Furthermore, due to the high levels of stellar activity, it can lead to increased stellar wind pressures that can strip away planetary atmospheres, particularly for close-in planets.

The RVs show evidence of an eccentric planet candidate ($e=0.66\pm0.09$) with a period of $427\pm7$~days and a minimum mass of $1.66\pm 0.26$~\Mjup. More data are needed to confirm this second planet, but this raises the possibility that TOI-4860\,b ended up in its small-periapsis orbit through planet-planet scattering or Kozai resonance with the outer planet, and then became circularised by tides.

\begin{acknowledgements}
We are grateful to the ESO/La Silla staff for their continuous support. We acknowledge funding from the European Research Council under the ERC Grant Agreement n. 337591-ExTrA. 

This paper includes data collected by the TESS mission. Funding for the TESS mission is provided by the NASA Explorer Program. We acknowledge the use of public TESS data from the pipelines at the TESS Science Office and at the TESS Science Processing Operations Center. Resources supporting this work were provided by the NASA High-End Computing (HEC) program through the NASA Advanced Supercomputing (NAS) Division at Ames Research Center for the production of the SPOC data products. 

Based on observations obtained at the Canada-France-Hawaii Telescope (CFHT) which is operated from the summit of Maunakea by the National Research Council of Canada, the Institut National des Sciences de l'Univers of the Centre National de la Recherche Scientifique of France, and the University of Hawaii. The observations at the Canada-France-Hawaii Telescope were performed with care and respect from the summit of Maunakea which is a significant cultural and historic site. Based on observations obtained with SPIRou, an international project led by Institut de Recherche en Astrophysique et Planétologie, Toulouse, France.

We thank the Swiss National Science Foundation (SNSF) and the Geneva University for their continuous support to our planet search programs. This work has been carried out within the framework of the National Centre of Competence in Research PlanetS supported by the Swiss National Science Foundation under grants 51NF40\_182901 and 51NF40\_205606. The authors acknowledge the financial support of the SNSF. 

A.J., R.B., and M.H.\ acknowledge support from ANID -- Millennium Science Initiative -- ICN12\_009. A.J.\ acknowledges additional support from FONDECYT project 1210718. R.B. acknowledges support from FONDECYT Project 11200751.

E.M. acknowledges funding from FAPEMIG under project number APQ-02493-22 and research productivity grant number 309829/2022-4 awarded by the CNPq, Brazil.

X.B., G.H., E.M., A.C., X.D., J.-F.D, T.F., P.F. and C.M. acknowledge funding from the French ANR under contract number ANR\-18\-CE31\-0019 (SPlaSH).

X.D., X.B., T.F., and A.C. acknowledge funding from the French National Research Agency in the framework of the Investissements d'Avenir program (ANR-15-IDEX-02), through the funding of the “Origin of Life” project of the Grenoble-Alpes University. 

E.G. acknowledges funding from NASA Award 80NSSC20K0957 (Exoplanets Research Program).

This work made use of \texttt{tpfplotter} by J. Lillo-Box (publicly available in www.github.com/jlillo/tpfplotter), which also made use of the python packages \texttt{astropy}, \texttt{lightkurve}, \texttt{matplotlib} and \texttt{numpy}. 

This research was carried out in part at the Jet Propulsion Laboratory, California Institute of Technology, under a contract with the National Aeronautics and Space Administration (NASA).
DR was supported by NASA under award number NNA16BD14C for NASA Academic Mission Services.

This work has made use of data from the European Space Agency (ESA) mission
{\it Gaia} (\url{https://www.cosmos.esa.int/gaia}), processed by the {\it Gaia}
Data Processing and Analysis Consortium (DPAC,
\url{https://www.cosmos.esa.int/web/gaia/dpac/consortium}). Funding for the DPAC
has been provided by national institutions, in particular, the institutions
participating in the {\it Gaia} Multilateral Agreement.

AC acknowledges support from CFisUC (UIDB/04564/2020 and UIDP/04564/2020), GRAVITY (PTDC/FIS-AST/7002/2020), and ENGAGE~SKA (POCI-01-0145-FEDER-022217), funded by COMPETE 2020 and FCT, Portugal. 
JFD acknowledges funding from the European Research Council (ERC) under the H2020 research \& innovation program (grant agreement \#740651 NewWorlds).

J.H.C.M. is supported in the form of a work contract funded by Fundação para a Ciência e Tecnologia (FCT) with the reference DL 57/2016/CP1364/CT0007; and also supported from by FCT - Fundação para a Ciência e a Tecnologia through national funds and by FEDER through COMPETE2020 - Programa Operacional Competitividade e Internacionalização by these grants: UIDB/04434/2020; UIDP/04434/2020. J.H.C.M. also acknowledges funding by the European Union (ERC, FIERCE, 101052347). Views and opinions expressed are however those of the author(s) only and do not necessarily reflect those of the European Union or the European Research Council. Neither the European Union nor the granting authority can be held responsible for them.

This research has made use of the Spanish Virtual Observatory (https://svo.cab.inta-csic.es) project funded by MCIN/AEI/10.13039/501100011033/ through grant PID2020-112949GB-I00.

The Pan-STARRS1 Surveys (PS1) and the PS1 public science archive have been made possible through contributions by the Institute for Astronomy, the University of Hawaii, the Pan-STARRS Project Office, the Max-Planck Society and its participating institutes, the Max Planck Institute for Astronomy, Heidelberg and the Max Planck Institute for Extraterrestrial Physics, Garching, The Johns Hopkins University, Durham University, the University of Edinburgh, the Queen's University Belfast, the Harvard-Smithsonian Center for Astrophysics, the Las Cumbres Observatory Global Telescope Network Incorporated, the National Central University of Taiwan, the Space Telescope Science Institute, the National Aeronautics and Space Administration under Grant No. NNX08AR22G issued through the Planetary Science Division of the NASA Science Mission Directorate, the National Science Foundation Grant No. AST-1238877, the University of Maryland, Eotvos Lorand University (ELTE), the Los Alamos National Laboratory, and the Gordon and Betty Moore Foundation.

This research was carried out in part at the Jet Propulsion Laboratory, California Institute of Technology, under a contract with the National Aeronautics and Space Administration (80NM0018D0004).

The Digitized Sky Surveys were produced at the Space Telescope Science Institute under U.S. Government grant NAG W-2166. The images of these surveys are based on photographic data obtained using the Oschin Schmidt Telescope on Palomar Mountain and the UK Schmidt Telescope.

\end{acknowledgements}

\bibliographystyle{aa}
\bibliography{refs.bib}

\begin{appendix} 
\FloatBarrier
\section{Additional figures and tables}

\begin{table}[htb]
\small
    \renewcommand{\arraystretch}{1.25}
    \setlength{\tabcolsep}{2pt}
\centering
\caption{Modelling of the SED.}\label{table.sed}
\begin{tabular}{lccc}
\hline
\hline
Parameter & & Prior & Posterior median   \\
&  & & and 68.3\% CI \\
\hline
Effective temperature, $T_{\mathrm{eff}}$ & [K]     & $N$(3191, 70)     & 3260$\pm50$ \\
Surface gravity, \logg\                   & [cgs]   & $U$(-0.5, 6.0)    & 5.4$^{+0.5}_{-0.7}$ \\
Metallicity, $[\rm{Fe/H}]$                & [dex]   & $N$(0.27, 0.12)  & $0.25\pm0.12$ \\
Distance                                  & [pc]    & $N$(80.42, 0.22)& $80.42 \pm 0.22$ \\
$E_{\mathrm{(B-V)}}$                      & [mag]   & $U$(0, 3)         & 0.039$^{+0.05}_{-0.029}$ \\
Jitter \textit{Gaia}                               & [mag]   & $U$(0, 1)         & 0.16$^{+0.21}_{-0.08}$ \\
Jitter 2MASS                              & [mag]   & $U$(0, 1)         & 0.027$^{+0.06}_{-0.020}$ \\
Jitter WISE                               & [mag]   & $U$(0, 1)         & 0.09$^{+0.19}_{-0.07}$ \\
Radius, $R_\star$                         & [\Rsun] & $U$(0, 100)       & 0.371$\pm0.010$ \\
Luminosity                                & [L$_\odot$] &               & 0.0139$\pm$ 0.0006 \smallskip\\
\hline
\end{tabular}
\tablefoot{$N$($\mu$,$\sigma$): Normal distribution prior with mean $\mu$, and standard deviation $\sigma$. $U$(l,u): Uniform distribution prior in the range [l, u].}
\end{table}

\begin{figure*}[h]
   \centering
   \includegraphics[width=1.0\textwidth]{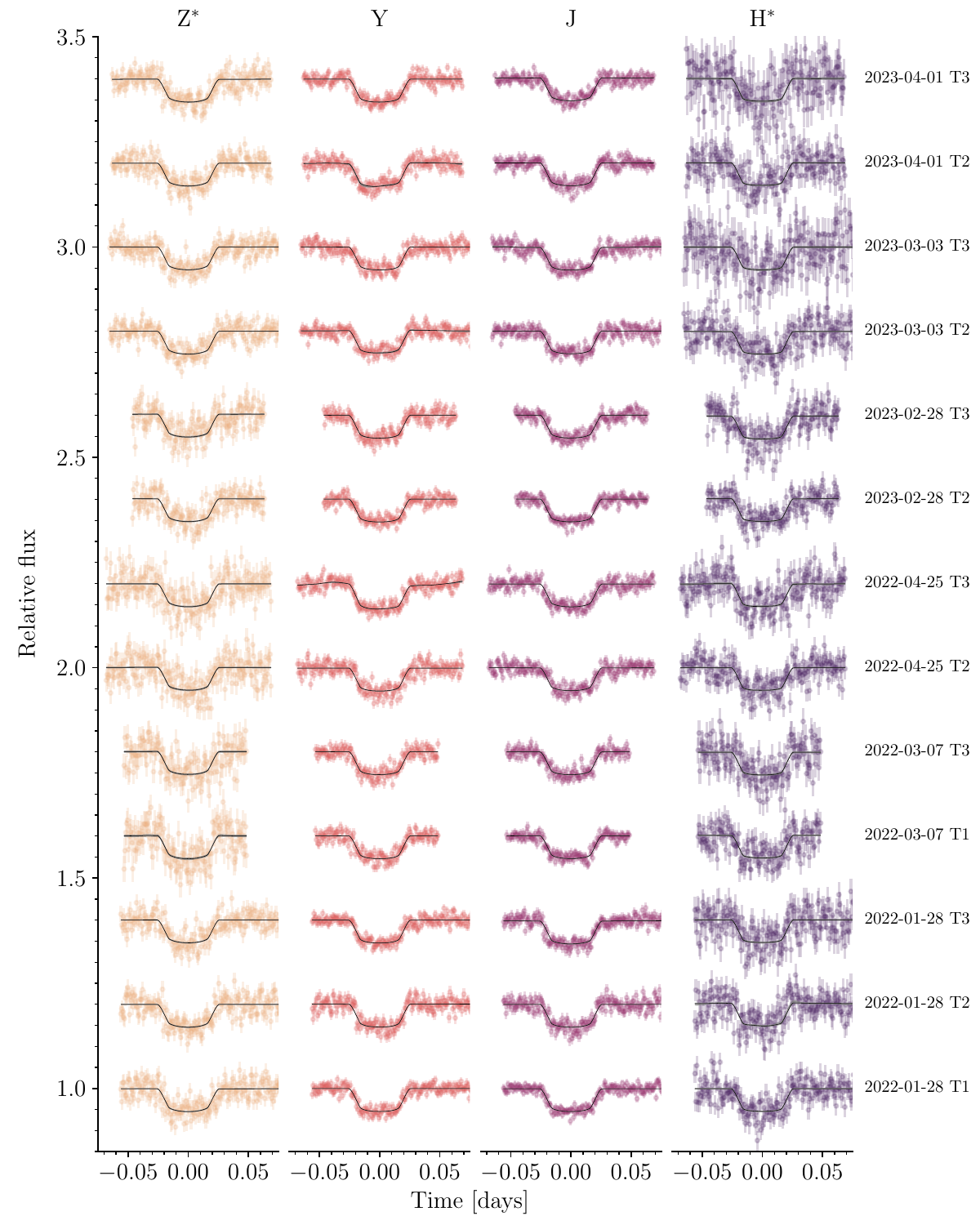}
      \caption{ExTrA transit photometry observations of TOI-4860\,b in the Z*, Y, J, and H* bands (error bars). Each line corresponds to a night (with format YYYY-MM-DD) and ExTrA telescope (labelled T1, T2, and T3) observation, which are offset vertically for clarity. For each transit, the median model (black line) and 68\% CI (grey band, barely visible) computed from 1000 random posterior samples are shown.}
      \label{fig:ExTrA_ZYJH}
\end{figure*}

\begin{figure}[h]
   \centering
   \includegraphics[width=0.5\textwidth]{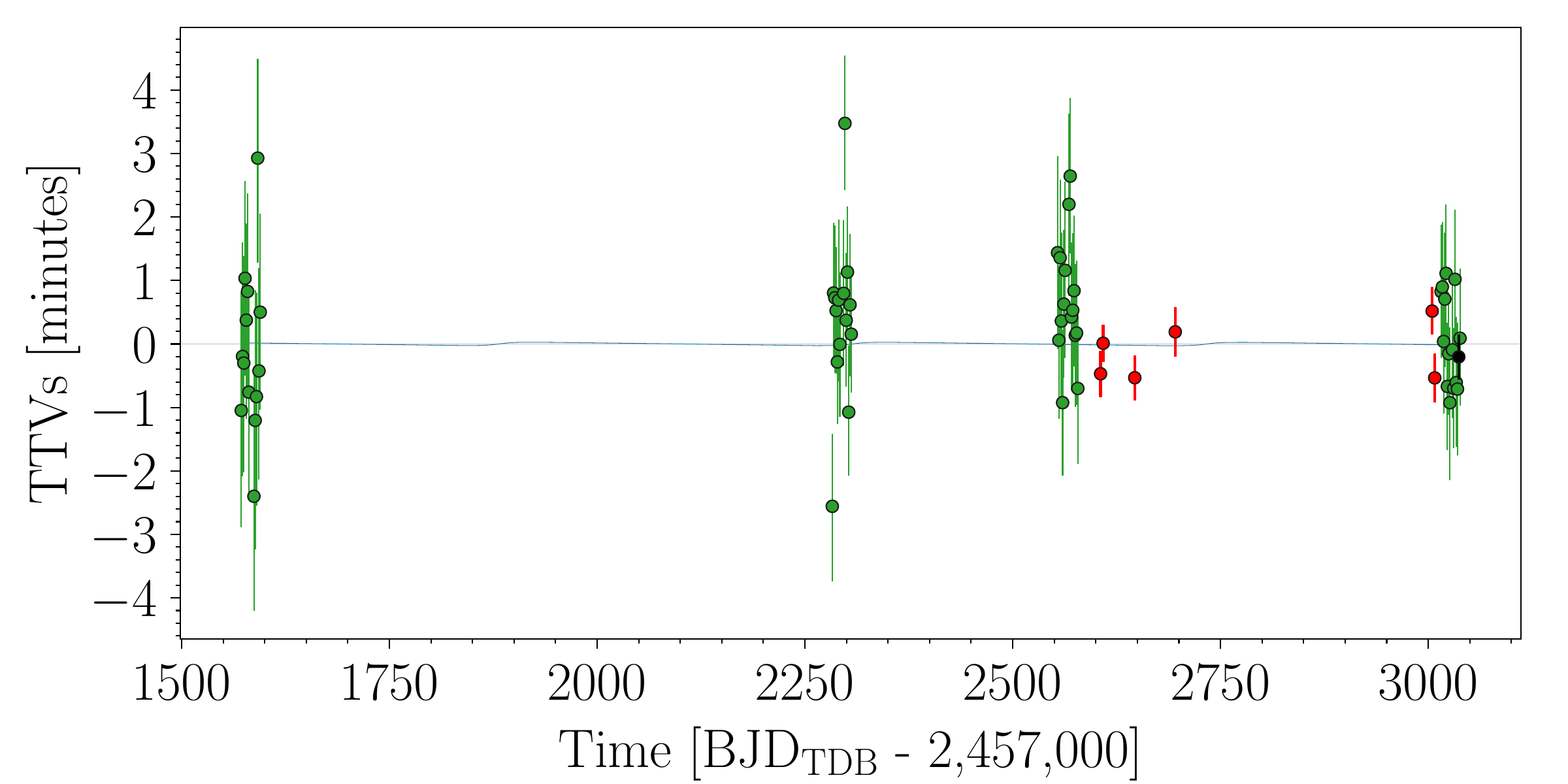}
      \caption{TTVs of TOI-4860\,b from TESS (green error bars), ExTrA (red error bars), and a transit observed simultaneously by TESS and ExTrA (black error bar). The blue curve is the expected TTVs due to the presence of the planet~c candidate.}
      \label{fig:TTV}
\end{figure}

\begin{figure}[h]
   \centering
   \includegraphics[width=0.48\textwidth]{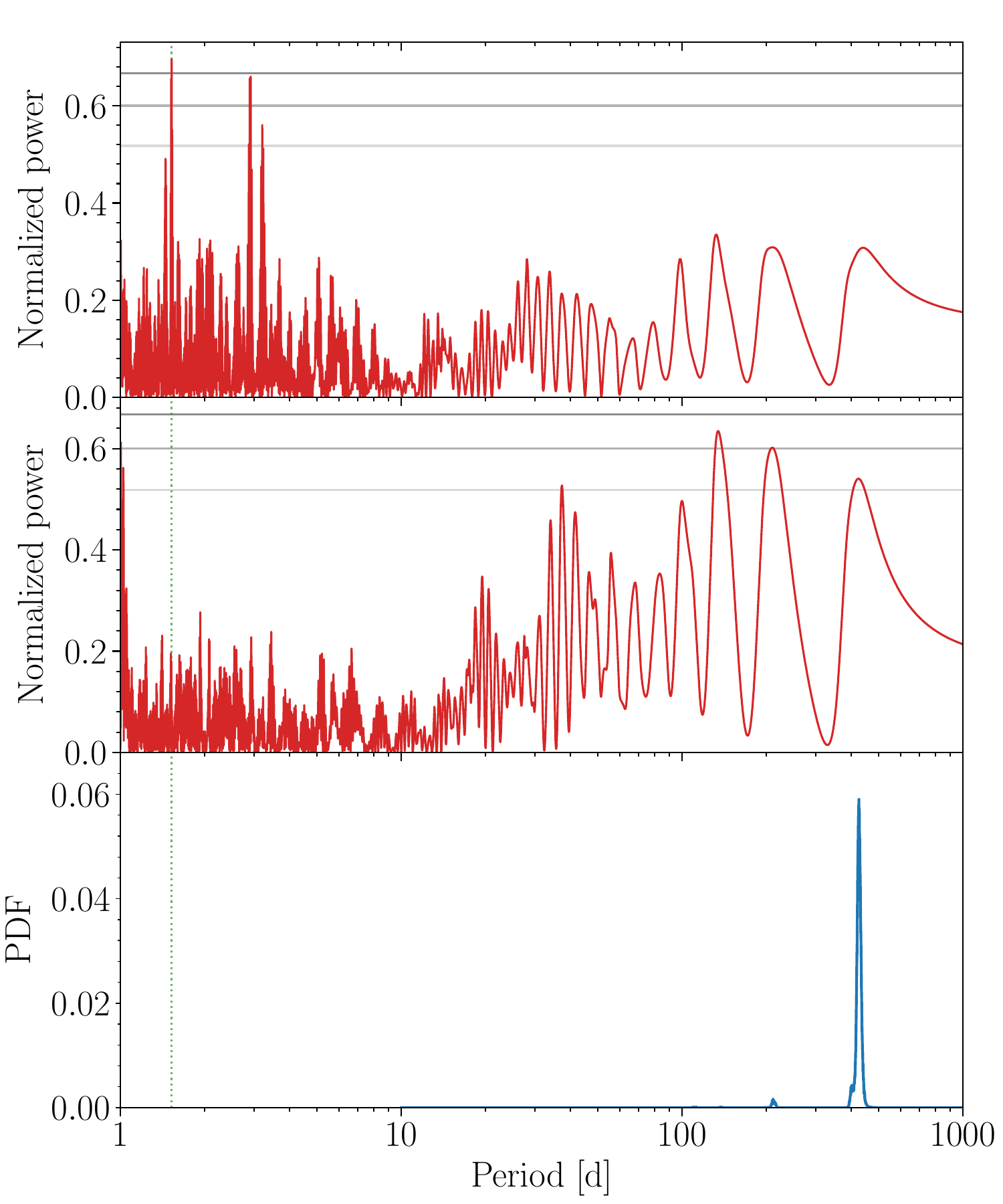}
      \caption{Periodogram of the SPIRou data. {\it Top panel:} Generalised Lomb–Scargle (GLS) periodogram \citep[][]{zechmeister2009} of the nightly averaged SPIRou velocities for TOI-4860 (red line). The horizontal lines represents 10, 1, and 0.1\% false-alarm levels, from bottom to top. The vertical dotted green line marks the period of the transiting planet. {\it Middle panel:} Idem after subtracting the Keplerian orbit of planet~b. Given that the planet~c candidate has a significant eccentricity, the power of the GLS periodogram is partially transferred to the harmonics of the orbital period, which leads to a reduced peak amplitude. {\it Bottom panel:} Marginal posterior distribution of the period of a second planet search with SPIRou and ESPRESSO (Sect.~\ref{section.RV}).}
      \label{fig:SPIRou_periodogram}
\end{figure}

\begin{table*}
\small
\caption{RV measurements.}\label{table.RVs}
\begin{tabular}{llrrl}
\hline
\hline
Time & RV &  $\pm1~\sigma$ & Exposure & Instrument\\
$[{\rm BJD_{TDB}}]$ & [\ms] & [\ms] & time [s] &\\
\hline
2459651.014632\tablefootmark{a} & 20672.79 & 35.56 & 184 & SPIRou \\
2459651.017018\tablefootmark{a} & 20728.73 & 38.07 & 184 & SPIRou \\
2459651.019405\tablefootmark{a} & 20727.25 & 38.52 & 184 & SPIRou \\
2459651.021792\tablefootmark{a} & 20783.75 & 41.44 & 184 & SPIRou \\
2459652.016441 & 20635.91 & 31.35 & 184 & SPIRou \\
2459652.018828 & 20655.03 & 32.13 & 184 & SPIRou \\
2459652.021215 & 20594.47 & 32.59 & 184 & SPIRou \\
2459652.023602 & 20609.83 & 33.11 & 184 & SPIRou \\
2459682.962344 & 20667.94 & 15.51 & 602 & SPIRou \\
2459682.969568 & 20689.38 & 15.49 & 602 & SPIRou \\
2459682.976857 & 20672.99 & 15.11 & 602 & SPIRou \\
2459682.984081 & 20675.06 & 15.54 & 602 & SPIRou \\
2459683.975195 & 20438.47 & 15.71 & 602 & SPIRou \\
2459683.982484 & 20553.25 & 15.54 & 602 & SPIRou \\
2459683.989773 & 20480.31 & 16.34 & 602 & SPIRou \\
2459683.996998 & 20506.34 & 17.13 & 602 & SPIRou \\
2459688.978921 & 20589.02 & 14.55 & 602 & SPIRou \\
2459688.986145 & 20617.96 & 15.26 & 602 & SPIRou \\
2459688.993370 & 20667.28 & 15.46 & 602 & SPIRou \\
2459689.000595 & 20677.68 & 15.52 & 602 & SPIRou \\
2459689.971959 & 20475.41 & 15.64 & 602 & SPIRou \\
2459689.979248 & 20550.51 & 15.97 & 602 & SPIRou \\
2459689.986473 & 20536.16 & 16.77 & 602 & SPIRou \\
2459689.993697 & 20539.39 & 16.53 & 602 & SPIRou \\
2459690.966227 & 20586.54 & 15.28 & 602 & SPIRou \\
2459690.973517 & 20577.52 & 15.71 & 602 & SPIRou \\
2459690.980745 & 20631.14 & 16.52 & 602 & SPIRou \\
2459690.988031 & 20561.25 & 16.73 & 602 & SPIRou \\
2459711.923333 & 20627.61 & 16.30 & 602 & SPIRou \\
2459711.930622 & 20583.55 & 17.47 & 602 & SPIRou \\
2459711.937910 & 20631.63 & 17.86 & 602 & SPIRou \\
2459711.945134 & 20662.24 & 17.92 & 602 & SPIRou \\
2459712.898086 & 20302.36 & 15.85 & 602 & SPIRou \\
2459712.905375 & 20438.74 & 18.02 & 602 & SPIRou \\
2459712.912599 & 20394.84 & 16.58 & 602 & SPIRou \\
2459712.919823 & 20455.80 & 17.72 & 602 & SPIRou \\
2459714.871905 & 20549.95 & 16.01 & 602 & SPIRou \\
2459714.879195 & 20642.84 & 16.56 & 602 & SPIRou \\
2459714.886419 & 20591.96 & 17.63 & 602 & SPIRou \\
2459714.893708 & 20547.20 & 18.88 & 602 & SPIRou \\
2459715.894581 & 20418.06 & 15.11 & 602 & SPIRou \\
2459715.901869 & 20457.61 & 15.26 & 602 & SPIRou \\
2459715.909158 & 20434.95 & 15.93 & 602 & SPIRou \\
2459715.916382 & 20499.60 & 17.42 & 602 & SPIRou \\
2459716.903709\tablefootmark{b} & 20550.12 & 13.69 & 602 & SPIRou \\
2459716.910997\tablefootmark{b} & 20534.58 & 14.51 & 602 & SPIRou \\
2459716.918222\tablefootmark{b} & 20486.82 & 14.86 & 602 & SPIRou \\
2459716.925446\tablefootmark{b} & 20534.29 & 15.74 & 602 & SPIRou \\
2459717.904838 & 20430.67 & 21.65 & 602 & SPIRou \\
2459717.912126 & 20535.44 & 25.12 & 602 & SPIRou \\
2459717.919415 & 20558.42 & 24.88 & 602 & SPIRou \\
2459717.926639 & 20623.72 & 27.34 & 602 & SPIRou \\
2459719.919542 & 20453.45 & 16.00 & 602 & SPIRou \\
2459719.926830 & 20539.12 & 17.01 & 602 & SPIRou \\
2459719.934054\tablefootmark{b} & 20542.62 & 27.27 & 602 & SPIRou \\
2459719.941278\tablefootmark{b} & 20477.03 & 23.83 & 602 & SPIRou \\
2459737.775387 & 20649.42 & 15.23 & 602 & SPIRou \\
2459737.782675 & 20682.05 & 15.45 & 602 & SPIRou \\
\hline
\end{tabular}
\begin{tabular}{llrrl}
\hline
\hline
Time & RV &  $\pm1~\sigma$ & Exposure & Instrument\\
$[{\rm BJD_{TDB}}]$ & [\ms] & [\ms] & time [s] &\\
\hline
2459737.789964 & 20635.10 & 16.42 & 602 & SPIRou \\
2459737.797252 & 20656.47 & 15.27 & 602 & SPIRou \\
2459741.833491 & 20510.06 & 14.71 & 602 & SPIRou \\
2459741.840779 & 20569.59 & 14.97 & 602 & SPIRou \\
2459741.848004 & 20518.22 & 15.84 & 602 & SPIRou \\
2459741.855228 & 20521.80 & 16.34 & 602 & SPIRou \\
2459974.822807 & 20519.66 & 1.18 & 2400 & ESPRESSO \\
2459975.790100\tablefootmark{b,c} & 20592.20 & 1.62 & 2400 & ESPRESSO \\
2459990.710841 & 20676.61 & 1.40 & 2400 & ESPRESSO \\
2459994.713548 & 20542.96 & 1.73 & 2400 & ESPRESSO \\
2459995.728387 & 20524.94 & 1.39 & 2400 & ESPRESSO \\
2459996.704515 & 20681.57 & 1.13 & 2400 & ESPRESSO \\
2459998.681241 & 20556.29 & 1.78 & 2400 & ESPRESSO \\
2460034.973275 & 20747.20 & 16.20 & 602 & SPIRou \\
2460034.980499 & 20691.66 & 15.74 & 602 & SPIRou \\
2460034.987723 & 20728.46 & 16.72 & 602 & SPIRou \\
2460034.994948 & 20750.87 & 16.75 & 602 & SPIRou \\
2460064.881853 & 20636.17 & 20.74 & 602 & SPIRou \\
2460064.889078 & 20628.63 & 38.55 & 602 & SPIRou \\
2460064.896367\tablefootmark{a,c} & 20688.17 & 50.19 & 602 & SPIRou \\
2460064.903591\tablefootmark{a,c} & 20764.02 & 60.23 & 602 & SPIRou \\
2460069.886522 & 20717.37 & 13.70 & 602 & SPIRou \\
2460069.893810 & 20705.09 & 14.03 & 602 & SPIRou \\
2460069.901034 & 20715.34 & 14.51 & 602 & SPIRou \\
2460069.908259 & 20698.83 & 14.95 & 602 & SPIRou \\
2460075.860806 & 20730.24 & 14.42 & 602 & SPIRou \\
2460075.868095 & 20768.85 & 13.83 & 602 & SPIRou \\
2460075.875383 & 20695.31 & 13.50 & 602 & SPIRou \\
2460075.882608 & 20707.40 & 13.94 & 602 & SPIRou \\
2460076.879995 & 20502.99 & 14.82 & 602 & SPIRou \\
2460076.887283 & 20542.48 & 16.61 & 602 & SPIRou \\
2460076.894508 & 20505.84 & 15.54 & 602 & SPIRou \\
2460076.901732 & 20556.08 & 15.46 & 602 & SPIRou \\
2460096.784419 & 20513.62 & 17.24 & 602 & SPIRou \\
2460096.791645 & 20606.75 & 16.41 & 602 & SPIRou \\
2460096.798874 & 20581.53 & 17.00 & 602 & SPIRou \\
2460096.806090\tablefootmark{a,c} & 21324.07 & 73.70 & 602 & SPIRou \\
2460098.813093 & 20731.23 & 25.57 & 602 & SPIRou \\
2460098.820377 & 20615.90 & 21.33 & 602 & SPIRou \\
2460098.827671 & 20714.14 & 22.61 & 602 & SPIRou \\
2460098.834895 & 20749.88 & 23.10 & 602 & SPIRou \\
2460102.846326 & 20522.33 & 13.25 & 602 & SPIRou \\
2460102.853615 & 20553.68 & 13.26 & 602 & SPIRou \\
2460102.860903 & 20551.93 & 13.50 & 602 & SPIRou \\
2460102.868127 & 20566.81 & 13.76 & 602 & SPIRou \\
2460124.740016 & 20611.59 & 16.83 & 602 & SPIRou \\
2460124.747305 & 20601.19 & 16.32 & 602 & SPIRou \\
2460124.754594 & 20615.81 & 16.28 & 602 & SPIRou \\
2460124.761820 & 20651.56 & 16.51 & 602 & SPIRou \\
2460128.751561 & 20552.08 & 18.63 & 602 & SPIRou \\
2460128.758785 & 20557.96 & 17.93 & 602 & SPIRou \\
2460128.766009 & 20520.20 & 18.43 & 602 & SPIRou \\
2460128.773233 & 20519.43 & 21.04 & 602 & SPIRou \\
2460129.750056 & 20450.99 & 24.34 & 602 & SPIRou \\
2460130.751754 & 20668.27 & 13.81 & 602 & SPIRou \\
2460130.759042 & 20649.19 & 13.81 & 602 & SPIRou \\
2460130.766331 & 20659.68 & 14.01 & 602 & SPIRou \\
2460130.773555 & 20691.74 & 14.74 & 602 & SPIRou \\
\hline
\end{tabular}
\tablefoot{\tablefoottext{a}{Flagged by the telluric correction algorithm.}\tablefoottext{b}{Observation during transit.}\tablefoottext{c}{Not used in the analysis.}}
\end{table*}

\begin{figure}[h]
   \centering
   \includegraphics[width=0.5\textwidth]{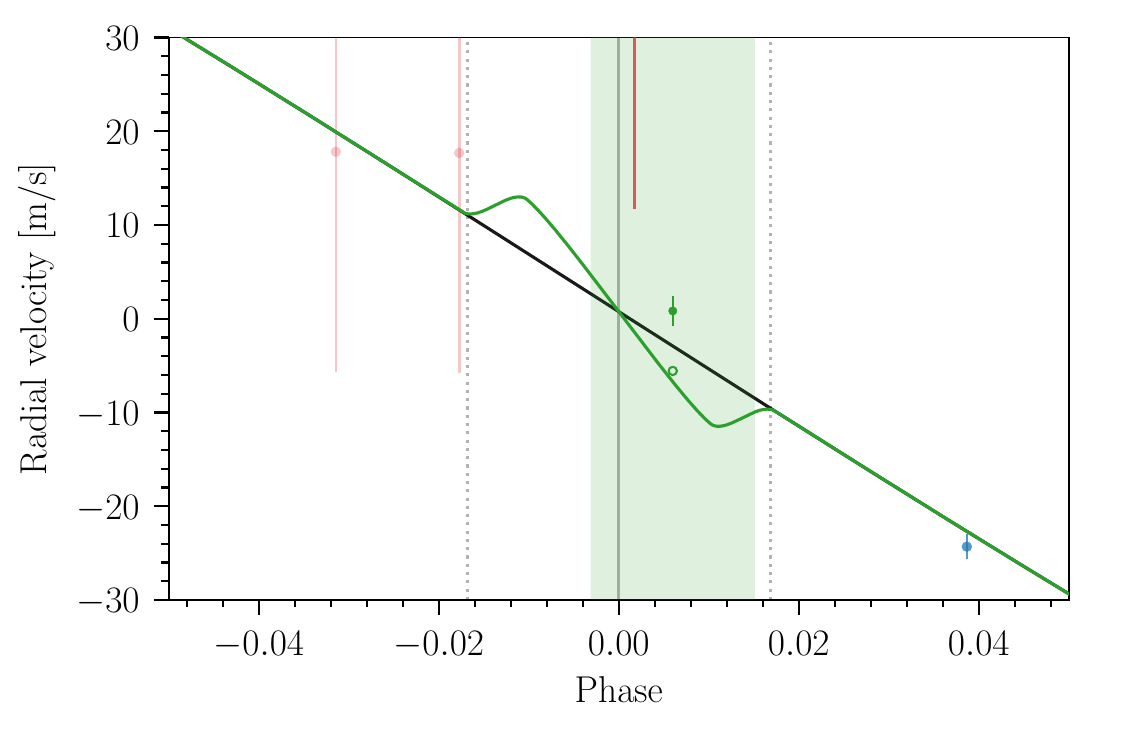}
      \caption{Same as the upper panel of Fig.~\ref{fig:RV_phase} but including the ESPRESSO observation during transit (green error bar), with its phase coverage highlighted in green. The transit centre is marked with a vertical solid grey line, and the ingress and egress are marked with dotted vertical grey lines. The green line is a model for the RVs that includes the Rossiter–McLaughlin effect for an aligned system \citep[computed with the \starry code;][]{luger2019}, and the green open circle is the integration of this model for the ESPRESSO observation during transit.}
      \label{fig:RM}
\end{figure}

\end{appendix}

\end{document}